\documentclass[onecolumn]{aastex701}

\usepackage{comment}
\usepackage{amsmath}

\pdfoutput=1

\received{26 October 2026}
\revised{-- --, 2026}
\accepted{-- --, 2026}

\newcommand{\kms}{ {km~s$^{-1}$}}
\newcommand{\kmss}{ {km~s$^{-1}$ }}

\newcommand{\dg}{$^{\circ}$ }

\hypersetup{linkcolor=blue,citecolor=blue,urlcolor=blue}

\shorttitle{Distant retrograde orbits for space weather forecasting}
\shortauthors{C. Möstl et al.}
\graphicspath{{./}{figures/}}

\begin{document}


\title{Calculation of distant retrograde orbits and their use in space weather forecasting}

\author[0000-0001-6868-4152]{Christian Möstl}
\affiliation{Austrian Space Weather Office, GeoSphere Austria, Graz, Austria}
\email{christian.moestl@geosphere.at}
\correspondingauthor{C. Möstl}
\email{christian.moestl@geosphere.at}

\author[0000-0001-9992-8471]{Emma E. Davies}
\affiliation{Austrian Space Weather Office, GeoSphere Austria, Graz, Austria}
\email{emma.davies@geosphere.at}

\author[0009-0004-8761-3789]{Eva Weiler}
\affiliation{Austrian Space Weather Office, GeoSphere Austria, Graz, Austria}
\affiliation{Institute of Physics, University of Graz, Graz, Austria}
\email{eva.weiler@geosphere.at}

\author[0000-0001-6590-3479]{Erika Palmerio}
\affiliation{Predictive Science Inc., San Diego, CA 92121, USA}
\email{epalmerio@predsci.com}

\author[0000-0002-1890-6156]{Noé Lugaz}
\affiliation{Space Science Center and Department of Physics and Astronomy, University of New Hampshire, Durham, NH 03824, USA}
\email{noe.lugaz@unh.edu}

\author[0009-0002-5200-644X]{Fabiana Camattari}
\affiliation{Department of Mathematics, University of Genoa, Genoa, Italy}
\affiliation{National Institute for Astrophysics, Astrophysical Observatory of Torino, Torino, Italy}
\email{fabiana.camattari@inaf.it}

\author[0000-0002-2559-2669]{Hannah T. Rüdisser}
\affiliation{Austrian Space Weather Office, GeoSphere Austria, Graz, Austria}
\affiliation{Institute of Physics, University of Graz, Graz, Austria}
\email{hannah.ruedisser@geosphere.at}

\author[0000-0003-1516-5441]{Ute V. Amerstorfer}
\affiliation{Austrian Space Weather Office, GeoSphere Austria, Graz, Austria}
\email{ute.amerstorfer@geosphere.at}




\begin{abstract}
There is current renewed interest in using Sun--Earth distant retrograde orbits (DROs) for a space weather forecasting mission, which would temporarily place spacecraft at a position near the Sun--Earth line, but closer to the Sun than the L1 point (such as the ESA HENON mission and ESA SHIELD concept). With in situ observations of the magnetic field on a DRO, the southward ($B_z < 0$) magnetic field of interplanetary coronal mass ejections (ICMEs), which is not routinely accessible remotely, could be measured several hours in advance. This $B_z < 0$ field is a decisive factor for forecasting geomagnetic storm intensity. We present an analysis of DROs at different sub-L1 distances for their efficacy for a space weather forecasting mission. An open-source numerical framework to generate DRO trajectories is presented. Our pre-calculated orbits in Heliocentric Earth Ecliptic coordinates can be introduced into numerical or empirical simulations of the ambient solar wind and ICMEs. We then analyze the corresponding forecast lead times, the resulting observational gaps depending on the number of spacecraft and distance of a DRO, and the ``cone of acceptance" of the maximum longitudinal separation of a spacecraft to the Sun--Earth line. We define different domains regarding radial evolution and longitudinal CME structure, and show the historical coverage of sub-L1 ICME observations with our ICMECAT catalog. An optimal mission configuration can be selected based on the characteristics of DROs presented here, depending on the requirements of such a mission deemed acceptable with respect to results on CME evolution and coherence.



\end{abstract}




\section{Introduction} \label{sec:intro}

Distant retrograde orbits \citep[DROs,][]{henon1969} are a family of heliocentric orbits that have recently come into focus for their potential use in space weather forecasting and for multipoint heliospheric solar wind observations. Spacecraft on a Sun--Earth DRO, during the portion of their orbit when they are positioned near the Sun--Earth line upstream of Earth, can measure parameters of a solar coronal mass ejection (CME) in situ, in particular their magnetic field orientation, and transmit the data to Earth in real time. This would allow the magnitude of the resulting geomagnetic storm to be forecast with a much improved lead time compared to data from the Sun--Earth L1 point. 

The advantages of DROs are that they are easier to reach than Venus-like heliocentric orbits \citep{ritter2015} and that they are stable over long periods \citep{henon1969, cyr2000, perozzi2017, cicalo2025henon}. Although in principle it is possible to extend their minimal distance along the Sun--Earth line to quite close to the Sun \citep{henon1969}, Venus at about 0.72 au poses a problem in the Solar System as it will perturb the orbits of spacecraft at any given DRO that crosses the distance of Venus's orbit at some point. However, the closer one places the DRO spacecraft to the Sun, the longer the potential warning lead time for a CME will be, at the cost of the orbit also becoming wider in longitude \citep{henon1969}. This means that there is a trade-off between forecasting lead time and correlation of the observations at the sub-L1 point and near Earth that needs to be studied for the optimal scientific and operational return of a space weather DRO mission.

The understanding of the 3D structure of CMEs and space weather forecasts could be greatly improved if several spacecraft were available at separations of up to 10--20\dg in heliospheric longitude and temporarily closer to the Sun--Earth line. These are known as sub-L1 monitors, and mostly case studies have so far revealed that this concept should be feasible for improved forecasts \citep[see][]{lindsay1999, cyr2000diamond, kubicka2016, laker2024, lugaz2025need, weiler2025, davies2026_solo}. The reason why sub-L1 monitoring should be possible is that CMEs consist of large-scale magnetic obstacles that propagate away from the Sun \citep[e.g.][]{burlaga1981, lepping1990, bothmer1998,mulligan1998,alhaddad2025structure}. While they undergo interactions with other structures and expand during propagation \citep[e.g.][]{manchester2017,luhmann_2020_review, temmer2021_review}, the magnetic flux ropes that form the core parts of CMEs tend to be stable enough over time and distance to be first measured somewhere closer to the Sun and later found with sufficient coherence in the near-Earth solar wind \citep[e.g.][]{good2018}. Large-scale magnetic flux ropes are thus the key property that make sub-L1 monitoring a viable method for CME forecasting, given the pivotal role of internal magnetic fields in regulating the space weather response at Earth. 

A major open question concerns how far away from Earth the sub-L1 monitor can be positioned, radially and longitudinally, to achieve the needed consistency in the observations with those at L1 for reliable space weather forecasts, and how large the gaps can be in between the spacecraft. Recent studies on multipoint CME observation have been made possible since 2020 with the current fleet of spacecraft including Solar Orbiter, Parker Solar Probe, BepiColombo, and STEREO-A in addition to L1 probes \citep[e.g.][]{moestl2022, regnault2024, banu2025, palmerio2025event4}. Power laws derived from ICME catalogs are highly useful as a general description of the ICME magnetic field and plasma properties evolution and expansion with heliocentric distance \citep[e.g.][]{leitner_2007, davies2021_juno, salman_2024_psp_cat, moestl2026}. Nevertheless, \cite{alhaddad2025structure} recently summarized the overall picture of CME flux ropes that has emerged from four decades of observations: a highly complex and irregular structure characterized by e.g.\ a kinked axis, a non-uniform cross-section, mixed open and closed magnetic field lines, and several ``bumps and lumps''. This raises the question of how far a sub-L1 monitor can be separated from the Sun--Earth line while still measuring structures that remain coherent with those later observed at L1, particularly given that individual features may be highly localized. \cite{owens2017} provided a theoretical basis for a limit of around 10\dg at 1~au, beyond which information may no longer propagate between two spatially separated observers. Consistent with this, \cite{banu2025} showed that the probability of jointly detecting a CME in situ at two spacecraft decreases below 50\% when their longitudinal separation exceeds 15$^{\circ}$. However, when both spacecraft do detect the same CME, the average magnetic field and plasma properties measured at the two locations remain comparable, indicating that a degree of coherence is retained even at these separations.

\cite{palmerio2025monitor} provided a historical summary of works that have employed serendipitous sub-L1 observations to predict CME magnetic fields and/or plasma quantities at 1~au in a hindcast fashion. While they all suggest that a sub-L1 monitor at any heliocentric distance and in a ``sufficient'' radial alignment with Earth should provide some degree of improvement to real-time forecasts, the available observations are too few and too widely distributed in space (between ${\sim}0.4$ and 0.96~au from the Sun, with longitudinal separations from Earth of up to 15$^{\circ}$) to draw systematic conclusions yet about the optimal placement of a sub-L1 monitor. \cite{lugaz2024_width} analyzed observations from STEREO-A as it approached the Sun--Earth line between 2020 and 2022 and found that the typical longitudinal width of magnetic obstacles is around 20\dg-- 30$^{\circ}$. \cite{weiler2026} analyzed STEREO-A observations during its passage across the Sun--Earth line in 2022--2024, covering separations of $\pm 15$\dg in heliospheric longitude relative to Earth. They found that separations of up to $\sim$0.05~au are likely insufficient to ensure that the monitor is sufficiently far upstream, as 25\% of the events considered impacted L1 before STEREO-A. The authors also found a potential longitudinal asymmetry, with greater (smaller) gains in lead time when STEREO-A was east (west) of the Sun--Earth line.

The first real-time demonstration of the potential of sub-L1 monitoring for improved space weather forecasts of CMEs was recently presented by \cite{davies2026_solo}. Their study employed Solar Orbiter as a far-upstream (${\sim}0.45$~au from the Sun) monitor with encouraging results for two separate CME events, with improved arrival time estimates in both cases while uncertainties remained reasonable. To assess the realistic benefits of a sub-L1 spacecraft or constellation, however, it is necessary to consider physically realizable orbits, as some configurations may be impractical or infeasible to implement. This shifts the focus toward attainable orbital configurations such as DROs. Historically, the ``Space Weather Diamond'' put forward by \cite{cyr2000} introduced DROs for space weather forecasting, with a four-spacecraft mission concept. Sub-L1 monitoring with solar sails has been studied with the Sunjammer concept in the 2010s \citep{eastwood_2015_sunjammer}, although the mission was never launched or operated. In the early 2020s, the MIIST mission concept using four spacecraft on DROs was put forward by \cite{lugaz2024ssr}, defining many of the main mission requirements---consisting of magnetic field, plasma, and solar energetic particle instruments---to ultimately improve space weather forecasting from in situ data upstream of L1. 

These are quite similar to the payload of HENON, a CubeSat that the European Space Agency (ESA) is currently building \citep[e.g.][]{cicalo2025henon}. It is planned to be launched in early 2027 on a rideshare with the PLATO mission, and transferred to a DRO with a minimum heliocentric distance of approximately 0.92~au. \cite{prete2026} performed numerical simulations with different initial conditions for the CME direction with EUHFORIA and concluded that at least within their modeled set-ups, HENON would provide a positive forecast lead time (i.e., the CME impacts first HENON and then Earth) when it resides within a ``key region" for upstream monitors, defined with a 60\dg opening angle as seen from Earth. HENON acts as a pathfinder for ESA SHIELD, which is another mission concept consisting of up to nine spacecraft currently studied in an early phase at ESA. However, we have not found in the literature a straightforward derivation and implementation for calculating DRO trajectories. This is necessary to enable their incorporation into empirical or numerical simulations of CME propagation \citep[e.g.][]{prete2026} and to evaluate their feasibility and practical utility. Ideally, a simple numerical scheme and physical units should be used so that the calculations are easily reproducible. Such a method would be highly useful as research on space weather forecasting methods needs access to use these orbits in order to optimize the scientific and operational return of missions such as ESA HENON and SHIELD. This paper attempts to close this gap.

With this work, we have three aims: (1) Show a straightforward method to numerically calculate DROs in physical units, to make them available for any type of heliospheric simulation. (2) Discuss their general characteristics to better understand their advantages and limitations with regard to space weather forecasts. (3) Place them within the context of known ICME in situ observations and define different physical domains relevant for upstream monitoring.


\section{Method} \label{sec:data}

There are no analytical solutions for DROs, hence they have to be numerically simulated, although initial conditions have been made available in the literature \citep{henon1969}. Our requirements for calculating these orbits are that these calculations should be easily accessible to everyone who needs to produce them on local machines, using a self-contained Python code package with as few dependencies as possible that is straightforward to install and run on any system. Physical units should be directly used in the simulations so that the results can be used without further steps, for introduction into numerical or empirical simulations of CMEs. The calculation should be done in a coordinate system that uses a fixed Sun--Earth line as the $x$-axis. With the ecliptic as the $x$--$y$ plane this is known as the Heliocentric Earth Ecliptic (HEE) coordinate system. 
This choice of coordinates makes the DRO easy to visualize, and many space weather forecasting models (e.g.\ Enlil, EUHFORIA, ELEvo) use the HEE or HEEQ system, with the Earth at position of around $x=1$ au on the right hand side of a visualization, and we make our plots consistent to this standard. We focus only on final orbits, and leave the transfer options to other authors \citep{cicalo2025henon} as this would need a proper treatment of full celestial dynamics, which is done with different software packages and is outside the scope of our study.

\subsection{Equations for the circular restricted three body problem}

Within the framework of the Circular Restricted Three-Body Problem (CR3BP), we present in Appendix~\ref{app:a} a derivation of the equation of motion from first principles, in two dimensions. The CR3BP assumes (1) a circular orbit of the Earth, (2) the mass of the Sun $m_1$ being larger compared to the mass of the Earth $m_2$ ($m_1 > m_2$), and (3) the spacecraft mass $m_3$ is negligible compared to both ($m_3 \ll m_1,m_2$). All three assumptions can be seen as well fulfilled in the system that we study. The final equations of motion in HEE coordinates are:

\begin{equation}
\label{eq:cr3bp1}
\begin{cases}
    \ddot{x} = 2 \omega \dot{y} + \omega^2 x - \dfrac{Gm_1}{p_1^3}(x-x_1) - \dfrac{Gm_2}{p_2^3}(x-x_2) \\
    \ddot{y} = -2\omega \dot{x} + \omega^2 y - \dfrac{Gm_1}{p_1^3}(y-y_1) - \dfrac{Gm_2}{p_2^3}(y-y_2),
\end{cases}
\end{equation}
with $(x,y)$ the position of the spacecraft, $(x_1,y_1)$ the position of the Sun, $(x_2,y_2)$ the position of the Earth, $p_1$ the distance between the spacecraft and the Sun, and $p_2$ the distance between the spacecraft and Earth. Our equations are stated in physical units, so after plugging them into a numerical solver, we can plot the resulting coordinates of the DRO spacecraft $(x,y)$ directly in a heliospheric coordinate system. With distances in units of km and velocities in km~s$^{-1}$ the gravitational constant is given as $G = 6.674 \times 10^{-20} \, \text{km}^3 \, \text{kg}^{-1} \, \text{s}^{-2}$, and $m_1=1.989 \times 10^{30} \, \text{kg}$ is the mass of the Sun and $m_2=5.972 \times 10^{24} \, \text{kg}$ is the mass of the Earth. The angular velocity of the Earth's rotation around the Sun is given as $\omega= 1.990987 \times 10^{-7}\, \text{rad~s}^{-1}$ (see Appendix~\ref{app:a} for details).

\subsection{Numerical implementation}

In the Python code that we make available as open source (see data availability section) we have made use of a simple numerical calculation scheme for the final differential Equations (\ref{eq:cr3bp1}).
We implement this solver in the python \textit{scipy} package with the \textit{solve\_ivp} function, using the DOP853 technique, a high-precision Runge--Kutta method. The computation for the trajectories for one year, for 11 different DROs as shown below, with a time resolution of 1 hour, takes less than a second on an office laptop.

Before the simulation of the DRO trajectory can be run, we need to find the initial conditions for the simulated spacecraft to reach the DRO orbits. This can in principle be done by trial and error, by placing the virtual spacecraft along a specific point on the orbit, which we chose as the coordinates $x,y=(r_{min},0)$, with $r_{min}$ being the minimum heliocentric distance of the DRO orbit on the Sun--Earth line, and then search for the value of the initial velocity component $v_{y;init}$ in the orthogonal $y$ direction that leads to a DRO trajectory that ends up at the starting point after one revolution around the Earth. Note that $v_{y;init}$ is always $>0$, because the spacecraft on a DRO moves in a clockwise direction around the $m_2$ body, which here is the Earth, as seen from ecliptic north. As the Earth moves in a counterclockwise direction around the Sun, the spacecraft's clockwise motion around the Earth gives rise to the term ``retrograde'' orbit.

\subsection{Finding DRO solutions}

\begin{figure*}
\centering
{\includegraphics[width=\textwidth]{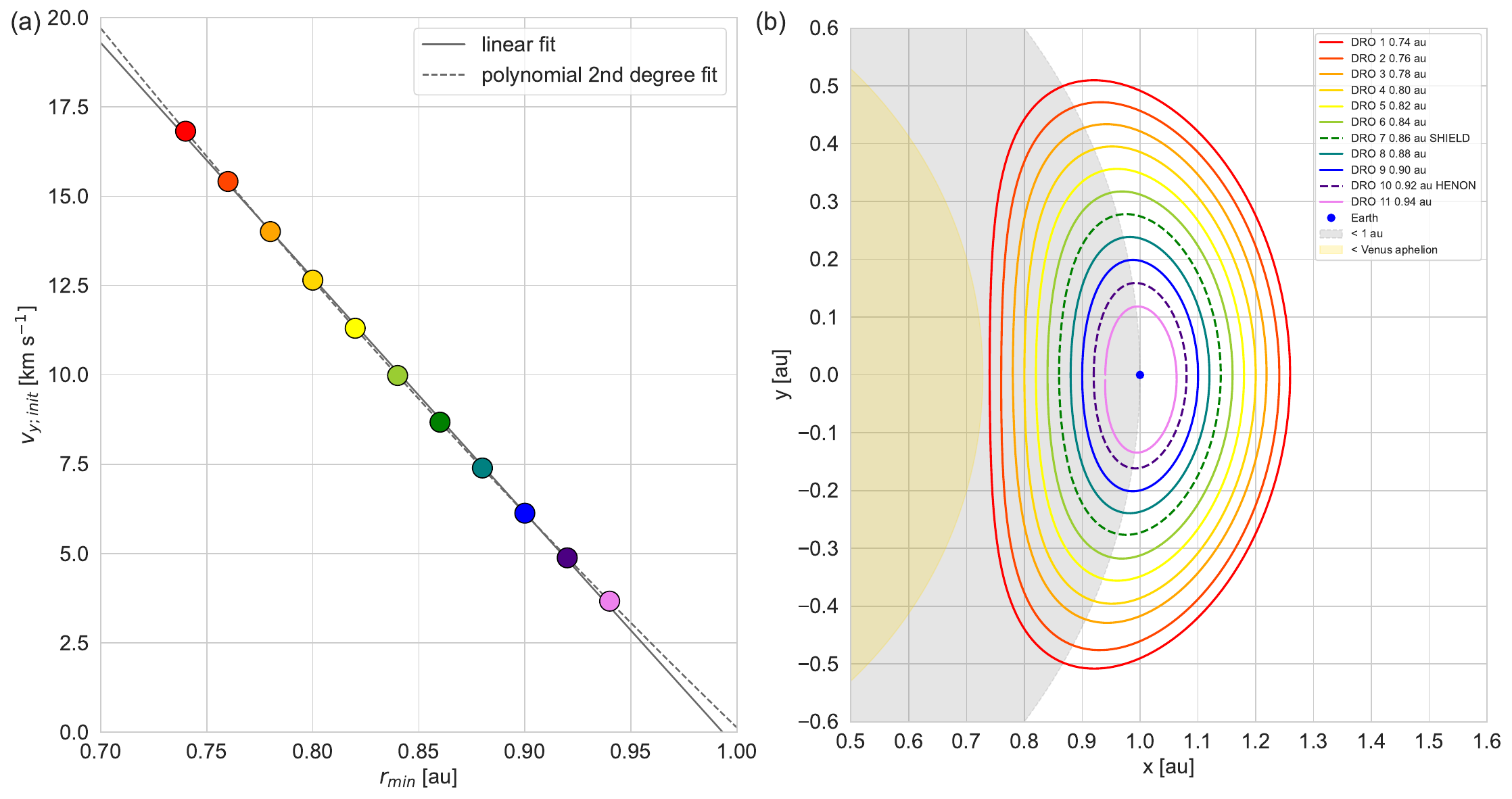}}
\caption{Finding distant retrograde orbits (DROs) with a numerical simulation of the CR3BP Equation~(\ref{eq:cr3bp1}). (a) We show the minimum distance $r_{min}$ along the Sun--Earth line of a given DRO versus the necessary initial velocity component $v_{y;init}$ in a 2D coordinate system for the virtual spacecraft to go into a DRO. Both a linear and a 2nd order polynomial fit are shown. (b) Solutions for 11 DROs with steps of 0.02 au, from 0.74 to 0.94 au, in an HEE coordinate system in which the Earth is at $x,y=(1,0)$ au. The area $< 1$ au is shaded grey, the area $<0.7282$ (Venus aphelion) is shaded in gold. The orbits representative for the missions SHIELD (0.86 au) and HENON (0.92 au) are shown as dashed lines.} 
\label{fig:dro_dmin_vinit}
\end{figure*}

We have devised a simple minimization scheme within our calculation setup to find the initial velocity $v_{y;init}$ at the point $x,y=(r_{min}, 0)$, with the velocity component in $x$ direction set to $v_{x;init}=0$. After one revolution of the synthetic spacecraft around the Earth, the end position should be as close as possible to the starting point, so the orbit is closed. In this way we find all numerical DRO solutions for any given $r_{min}$. 

Figure~\ref{fig:dro_dmin_vinit}a shows the resulting almost linear relationship between $v_{y;init}$ and $r_{min}$. A linear fit as an approximation is:
\begin{equation}
v_{y;init}\, \text{[km~s}^{-1}] = -65.8019 \: r_{min} \text{ [au] }+ 65.3602
\end{equation}
and a polynomial on the order of 2nd degree is only marginally closer to the data points than a linear fit. 

Figure~\ref{fig:dro_dmin_vinit}b visualizes 11 representative orbit solutions we found with this method, in a cartesian coordinate system in Heliocentric Earth Ecliptic (HEE) coordinates. We will discuss all results in the context of 11 orbits in the range of $0.74 \leq r_{min} \leq 0.94$, in steps of 0.02 au. For less cluttered visualizations, we will further show the 4 orbits consisting of $r_{min}=0.74$ au (innermost solution), 0.80 au, 0.86 au (representative for the SHIELD mission), and 0.92 au (for the HENON mission).

We choose this range for the reasons that for distances $> 0.94$~au, there is too little benefit in terms of lead time of a DRO compared to the L1 point \citep{weiler2026}, which is situated 0.01~au upstream of Earth or 0.99~au from the Sun. For distances $r_{min} <0.74$~au we assume that the orbit of Venus at around 0.72 au may start to influence the DRO over time. Below $r_{min}< 0.80$~au the DROs become, as we will see below, very wide in the longitudinal direction, and thus the number of spacecraft that would be needed to provide dense enough coverage of the space near the Sun--Earth line is very likely unrealistic. This also makes it unnecessary to identify the minimum distance from Venus at which a DRO remains unperturbed, hence we do not discuss this issue further here.


\section{Results}

\begin{figure*}
\centering
{\includegraphics[width=\textwidth]{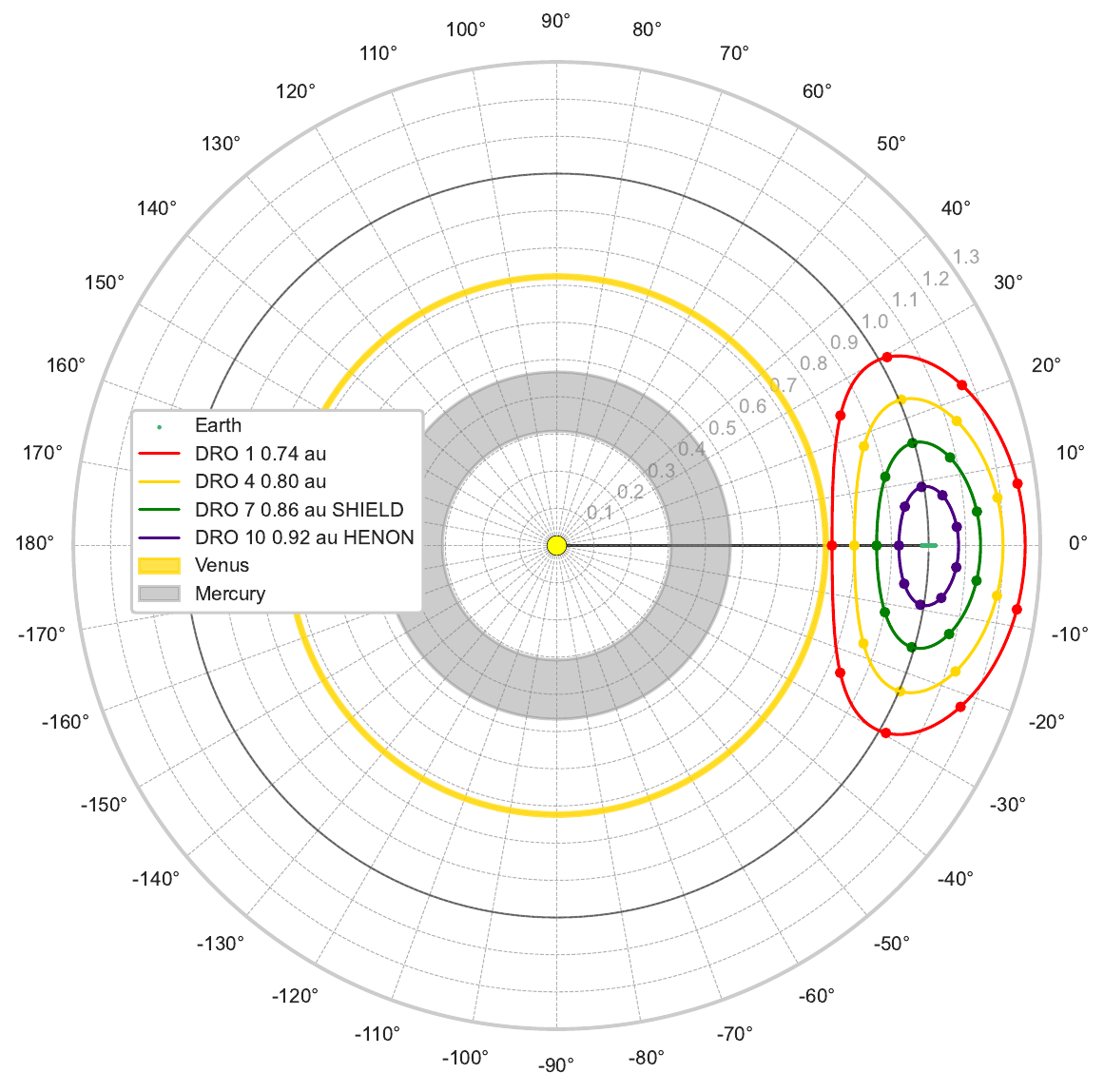}}
\caption{Distant retrograde orbit configuration in the inner solar system context for four different minimum heliocentric distances $d_{x;min}$=(0.74, 0.80, 0.86, 0.92 au). The shaded areas give the planetary orbit extension from perihelion to aphelion of Mercury, Venus and Earth, in an Heliocentric Earth Ecliptic (HEE) coordinate system. Each dot on a DRO visualizes a spacecraft position, with 9 spacecraft equidistantly distributed on each DRO. This figure is available as an animation in the online journal.
} 
\label{fig:dro_planets}
\end{figure*}

Figure~\ref{fig:dro_planets} shows the four main DROs (0.74 au, 0.80 au, 0.86 au, 0.92 au) compared to the orbit of Mercury and Venus in a view of the inner solar system. For each orbit, we added 9 spacecraft spread equidistant in time over the full DRO, which will later play a role when we assess how large the gaps are in between the spacecraft in heliospheric longitude. We now discuss some main characteristics of DROs.

\subsection{Size of orbit} \label{subsec:size}

\begin{figure*}
\centering
{\includegraphics[width=15 cm]{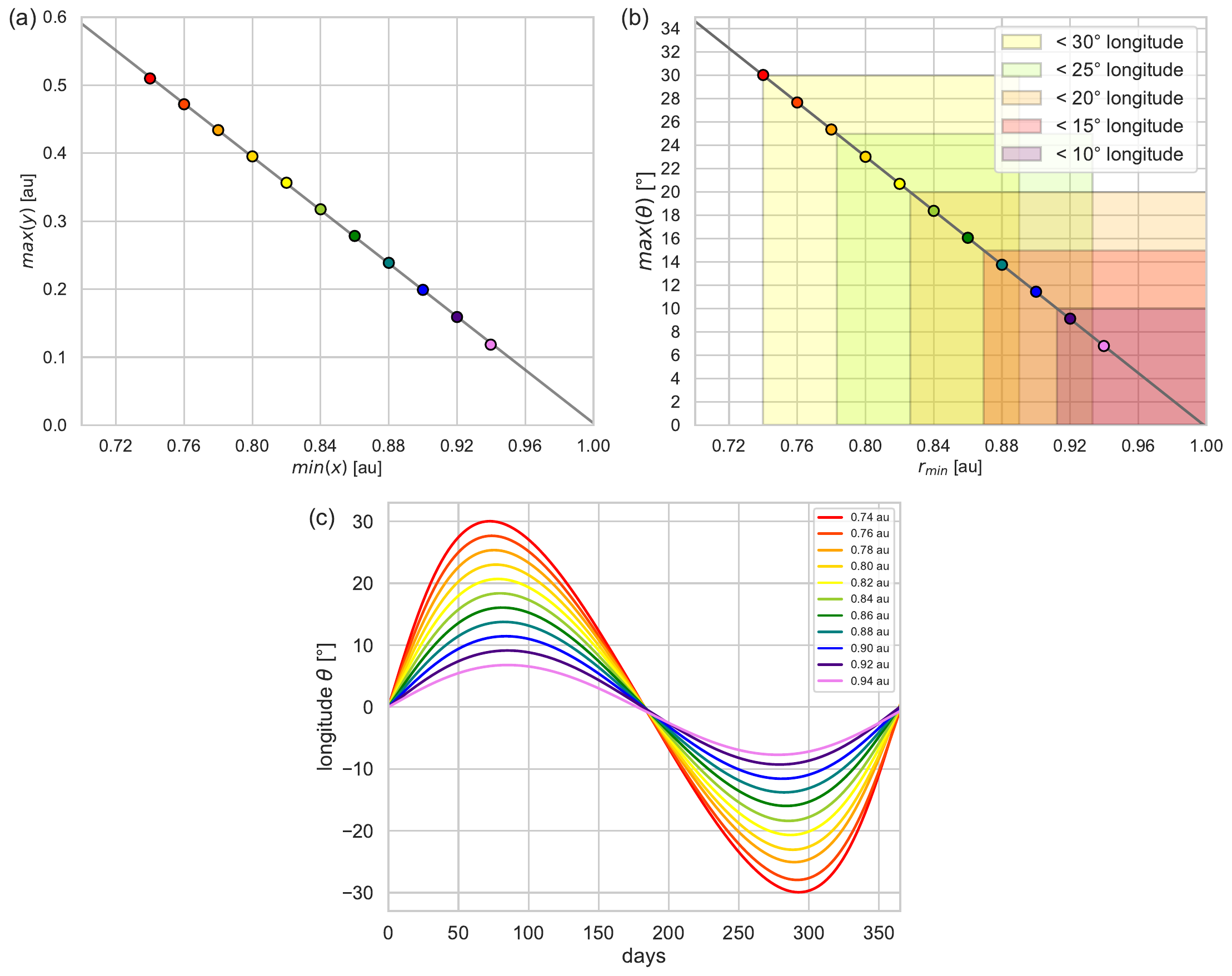}}
\caption{Characteristics of DROs. (a) The minimum distance along the Sun--Earth line of a given DRO $min(x)=r_{min}$ and the farthest extension in HEE $max(y)$ in au. (b) The minimum distance along the Sun--Earth line $r_{min}$ versus the maximum HEE longitude extension $max(\theta)$ in degrees. Domains for which the DRO is always inside 10° to 30° longitude are shown as shaded rectangles (see legend). (c) A plot of the HEE longitude along the orbit versus time demonstrates that the orbital periodicity for all DROs is one year, as given by the Earth's rotation around the Sun.
}
\label{fig:dro_dmin_max_long_period}
\end{figure*}
Figure~\ref{fig:dro_dmin_max_long_period}a shows the important relationship between the DRO extension in radial distance along the Sun--Earth line, and the corresponding maximum extension in the orthogonal direction (y coordinate in HEE). The ratio of the orthogonal to radial extent is very close to a factor of 2 ($\approx 1.98$), which can be seen as an easy-to-remember rule of thumb. This means that there is a clear tradeoff between going closer to the Sun and having excursions farther away from the Sun--Earth line, which is independent of the exact DRO and has a factor of $\sim$2 ratio. As further discussed below, this tradeoff can be mitigated by deploying a constellation with more spacecraft. Another essential feature is the symmetry in the minimum heliocentric distance along the Sun--Earth line with the maximum heliocentric distance of the orbit, and the distance of these two points from Earth is similar, for all DROs.

The shape of a DRO can be thought of as ellipse-like or bean-shaped, with the ``front'' part facing the Sun being flatter than the ``back'' behind the Earth. These shapes, however, are not self-similar, and each solution for a given $r_{min}$ is slightly different. In particular, the ``flattening'' of the front part becomes more pronounced for DROs with an $r_{min}$ closer to the Sun. Interestingly, this intrinsic feature is advantageous to space weather forecasting, as the upstream distance remains closer to the Sun, even if the spacecraft is further away from the Sun--Earth line, the more upstream a DRO is situated. 

In Figure~\ref{fig:dro_dmin_max_long_period}b, we show the maximum longitude extension of a given DRO against $r_{min}$, with the data also given in Table~\ref{tab:1dmin_maxtheta} for reference. For example, if the specification for a DRO mission is that the maximum extension in longitude is 15°, determined e.g.\ by a limit on CME coherence, then the $r_{min}$ of the DRO cannot be ${<}0.8692$ au. We show a linear fit to the data points for each of the 11 DROs in Figure~\ref{fig:dro_dmin_max_long_period}b, that results in
\begin{equation}
max(\theta) \text{[°]} =  -115.9720 \: r_{min} \text{[au]} +115.8077  .
\end{equation}
This main characteristic of DROs---that the extension in longitude becomes wider for orbits that reach smaller heliocentric distances---is a very important constraint that sets limits on how close to the Sun it is useful to place a spacecraft on these types of orbits. While it is highly desirable to get closer to the Sun, which gives us longer forecast lead times, the orbits become wider so that more spacecraft are needed to maintain sufficiently small angular separations between them.


\begin{table}[h]
\centering
\begin{tabular}{cc|cc|cc|cc|cc}
\toprule
$max(\theta)$ [°] & $r_{min} [au]$ & $max(\theta)$ [°] & $r_{min}$ [au] & $max(\theta)$ [°] & $r_{min}$ [au] & $max(\theta)$ [°] & $r_{min}$ [au] & $max(\theta)$ [°] & $r_{min}$ [au]  \\
\hline
5  & 0.9554 & 11 & 0.9037 & 17 & 0.8519 & 23 & 0.8002 & 29 & 0.7485 \\
6  & 0.9468 & 12 & 0.8951 & 18 & 0.8433 & 24 & 0.7916 & 30 & 0.7399 \\
7  & 0.9382 & 13 & 0.8864 & 19 & 0.8347 & 25 & 0.7830 & 31 & 0.7312 \\
8  & 0.9296 & 14 & 0.8778 & 20 & 0.8261 & 26 & 0.7743 & 32 & 0.7226 \\
9  & 0.9209 & 15 & 0.8692 & 21 & 0.8175 & 27 & 0.7657 & 33 & 0.7140 \\
10 & 0.9123 & 16 & 0.8606 & 22 & 0.8088 & 28 & 0.7571 & 34 & 0.7054 \\
\end{tabular}
\caption{Maximum extension in heliospheric longitude $max(\theta)$ of a distant retrograde orbit (DRO) and the corresponding minimum heliocentric distance $r_{min}$ along the Sun--Earth line. This corresponds to the data shown in Figure~\ref{fig:dro_dmin_max_long_period}b.}
\label{tab:1dmin_maxtheta}
\end{table}

\subsection{Orbital period}
In Figure~\ref{fig:dro_dmin_max_long_period}c, the longitude $\theta$ of all 11 DROs (covering the range between 0.74 and 0.94~au in increments of 0.02~au) is plotted against time. This demonstrates another issue of major importance concerning the characteristics of DROs. All orbits take  one year to complete one apparent revolution around the Earth, regardless of heliocentric distance. This means that the orbital period is tied to the second body of the CR3BP, which in our case is Earth. Note the asymmetry in the longitude versus time plot, which is a consequence of the slightly asymmetric ``bean" shape that is ``flatter" during the crossing of the line between the Sun and the planet, as seen in Figure~\ref{fig:dro_dmin_vinit}b. This asymmetry becomes more pronounced with increasing heliocentric distance, as mentioned in Section~\ref{subsec:size}. The reason for this is the higher orbital speed during the Sun--Earth line crossing, when the spacecraft is closest to the Sun, resulting in the slightly asymmetric ``bean'' shape.


One major consequence of the fact that all orbits have a period of one year is that, in principle, adding spacecraft to an existing constellation is possible, although the transfers would need to be timed correctly. For example, for an aging spacecraft fleet on a DRO of 0.86 au, some spacecraft that are no longer operational could be replaced by deploying novel spacecraft into the correct ``phase'' with respect to the existing assets, regardless of distance. Even though all spacecraft in a DRO constellation are on different \textit{heliocentric} orbits, in a HEE system these orbits can be thought of being in a different ``phase'' with each other.

If it is deemed that another launch into a 0.86 au orbit is too expensive, the replacement spacecraft could also be put into an orbit at e.g.\ 0.92 au. As long as the initial ``phase'' placement with the ``older" spacecraft is correct, their phase with respect to each other will stay similar as the orbital period of one year is independent of distance. Alternatively, a second fleet of DRO spacecraft could be placed at different distances, for added scientific and operational value concerning the differences in radial and longitudinal distance to the first fleet. If the ``mesh'' of the first fleet is deemed too coarse, in the sense that the distances between spacecraft are too wide, another fleet could be placed into any other DRO, thereby making the mesh denser, without needing to adjust the orbit in time.

\subsection{Cone of acceptance and key region}

\cite{prete2026} discuss the concept of the ``key region", which is a characteristic of every upstream monitoring mission. They defined this as an upstream region extending in a 60° angle as seen from Earth. However, this concept can be refined based on the heliospheric longitude difference that is acceptable. We introduce hereby the ``cone of acceptance'' $\Lambda$, which is the angle from the Sun--Earth line to a given line that delimits the region useful for employing the upstream in situ data for making predictions at the Earth's location.

Table~\ref{tab:r_lambda1} shows values for $\Lambda$, depending on each DRO defined its minimum heliocentric distance along the Sun--Earth line $r_{min}$, and the corresponding heliocentric distance of the point along the DRO at the point of each $\Lambda$ value. The main takeaway here is that we actually need \textit{two} parameters to define a key upstream region suitable for forecasting: the cone of acceptance $\Lambda$ (in degrees, HEE), and a limit on the acceptable upstream heliocentric distance along the DRO, which we call $r_{limit}$ (in au). For all suitable upstream measurements for forecasts, $r_{min} < r < r_{limit}$. Both values need to be determined from CME and shock physics and observations; a future meta study of available research results and future studies should settle $\Lambda$ and $r_{limit}$, but reasonable values seem to cluster around $\Lambda=5-15$° and $r_{limit}=0.85-0.95$ au.

For example, for a SHIELD-like mission configuration with $r_{min}=0.86$~au, a cone of acceptance of $\Lambda=10$° leads to an $r_{limit}$ of 0.89 au. As noted before, it is important to point out that the shape of DROs helps in this respect, because the sunward part of the ``bean'' is flatter than the rear, and so the spacecraft does stray not too far away from $r_{min}$ when moving a few degrees away from the Sun--Earth line, but then at some point more quickly turn towards 1 au. Giving attention to $r_{limit}$ is particularly important to 
ensure that upstream distances are used for which $r$ does not deviate too much from $r_{min}$. This is because the forecast lead times derived from observations at the edge of the upstream monitoring domain given by $\Lambda$ should not be too much lower than those derived from the distance $r_{min}$. We do not want to ``lose'' too much prediction lead time when a spacecraft situated near the longitude of $\Lambda$ is used for forecasts.


\begin{table}[h]
\centering
\begin{tabular}{c|ccccccc}
\toprule
 & 5.0° & 7.5° & 10.0° & 12.5° & 15.0° & 17.5° & 20.0° \\
\hline
0.74 au & 0.7431 & 0.7471 & 0.7529 & 0.7605 & 0.7704 & 0.7827 & 0.7982 \\
0.76 au & 0.7635 & 0.7679 & 0.7742 & 0.7828 & 0.7939 & 0.8081 & 0.8264 \\
0.78 au & 0.7838 & 0.7887 & 0.7959 & 0.8056 & 0.8182 & 0.8348 & 0.8569 \\
0.80 au & 0.8043 & 0.8098 & 0.8179 & 0.8290 & 0.8440 & 0.8641 & 0.8933 \\
0.82 au & 0.8249 & 0.8312 & 0.8404 & 0.8535 & 0.8716 & 0.8979 & 0.9459 \\
0.84 au & 0.8455 & 0.8529 & 0.8639 & 0.8797 & 0.9033 & 0.9452 & -- \\
0.86 au & 0.8665 & 0.8751 & 0.8885 & 0.9090 & 0.9454 & -- & -- \\
0.88 au & 0.8877 & 0.8983 & 0.9157 & 0.9468 & -- & -- & -- \\
0.90 au & 0.9096 & 0.9234 & 0.9494 & -- & -- & -- & -- \\
0.92 au & 0.9326 & 0.9533 & -- & -- & -- & -- & -- \\
0.94 au & 0.9590 & -- & -- & -- & -- & -- & -- \\
\end{tabular}%
\caption{This table demonstrates the increase in heliocentric distance on each distant retrograde orbit when moving away from the Sun--Earth line. Each table entry gives the heliocentric distance $r_{limit}$ on a DRO at the point of a given longitude value for the cone of acceptance $\Lambda$ (columns), for each of the eleven DROs defined by $r_{min}$ (rows). 
} \label{tab:r_lambda1}
\end{table}

\subsection{Forecast lead times}

\begin{figure*}
\centering
{\includegraphics[width=\textwidth]{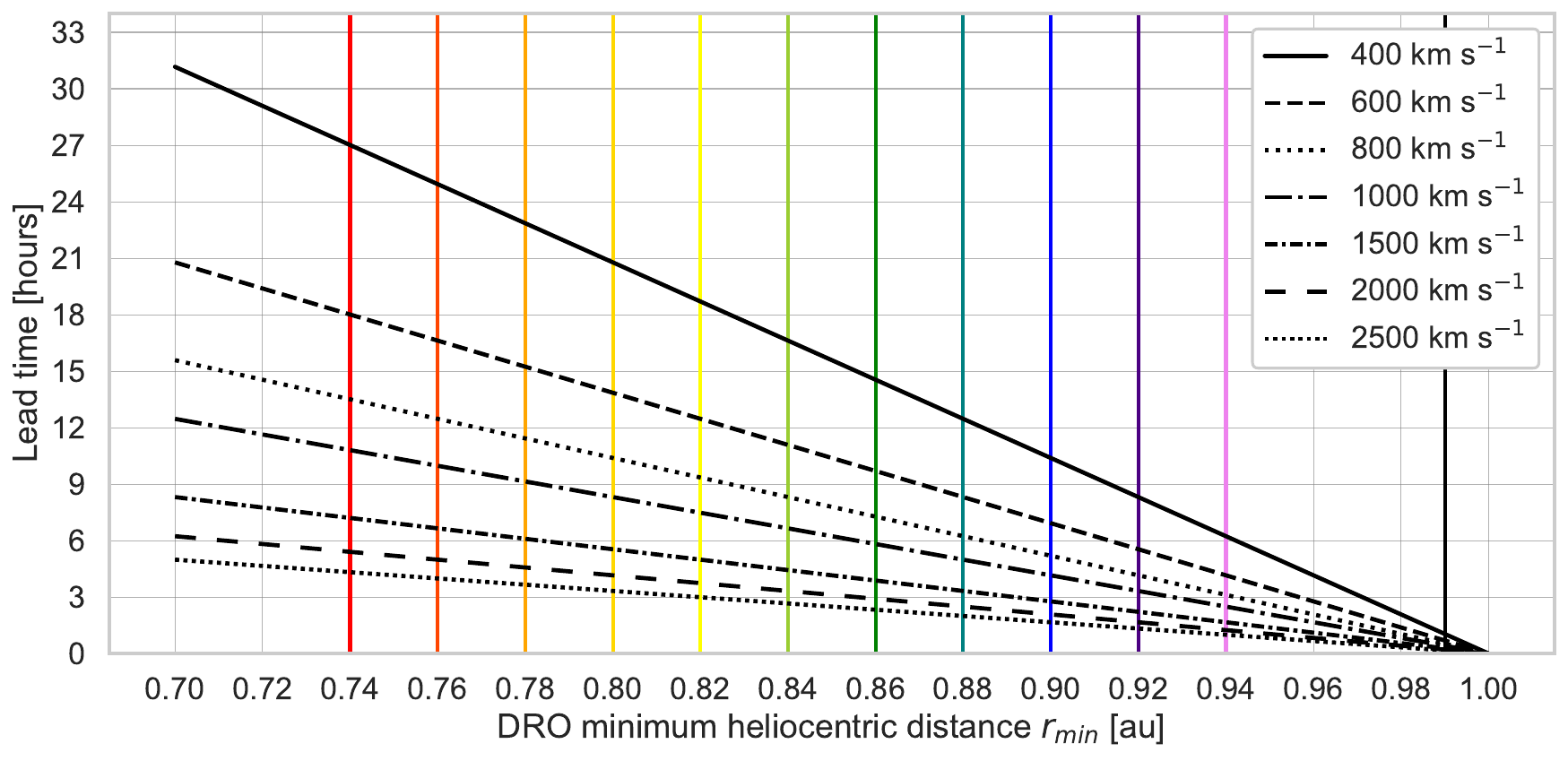}}
\caption{Prediction lead times for spacecraft on DROs. We assume propagation with constant speed along the Sun--Earth line, from the sub-L1 monitor position to 1 au.  Different CME speeds are denoted by different line styles. Vertical colored lines indicate the 11 DRO minimum distances $r_{min}$ to the Sun along the Sun--Earth line (0.74--0.94 au), with the color code similar to previous figures. The black vertical line on the very right is the distance of L1 at 0.99 au.} 
\label{fig:lead}
\end{figure*}

Figure~\ref{fig:lead} shows the lead times for different CMEs with constant speed from the DRO to 1~au, assuming radial propagation along the Sun--Earth line with a constant speed for simplicity. In comparison to the lead times with in situ data from the L1 point, a CME with a speed of 400~\kmss takes about 1 hour to reach the Earth's magnetopause, and for higher speeds all lead times are $< 1$~hour, with only 10--15 minutes for the fastest CMEs, resulting in a nowcast rather than a forecast. Table~\ref{tab:leadtime} summarizes the results, for each DRO configuration and CME speed. For a CME traveling at 400~\kms, as a rule of thumb, each 0.01 au further upstream adds roughly one hour of forecast lead time. For a SHIELD-like DRO at 0.86 au, the lead times range from 14.5 hours for slow CMEs to 2.3 hours for an extremely fast CME (2500 \kms). This speed is actually higher than the highest speed observed in situ near 1 au \citep{liu2014}, thus setting a strong lower limit on reachable forecast lead times. Recently, \cite{horne2026_maystorm_uk} state a forecast lead time of 2--3 hours is needed for making useful actionable forecasts.

\begin{table}[h]
\centering
\resizebox{\textwidth}{!}{%
\begin{tabular}{c|ccccccc}
\toprule
 & 400 \kms & 600 \kms & 800 \kms & 1000 \kms & 1500 \kms & 2000 \kms & 2500 \kms \\
\hline
0.74 au & 27.0 & 18.0 & 13.5 & 10.8 & 7.2 & 5.4 & 4.3 \\
0.76 au & 24.9 & 16.6 & 12.5 & 10.0 & 6.6 & 5.0 & 4.0 \\
0.78 au & 22.9 & 15.2 & 11.4 & 9.1  & 6.1 & 4.6 & 3.7 \\
0.80 au & 20.8 & 13.9 & 10.4 & 8.3  & 5.5 & 4.2 & 3.3 \\
0.82 au & 18.7 & 12.5 & 9.3  & 7.5  & 5.0 & 3.7 & 3.0 \\
0.84 au & 16.6 & 11.1 & 8.3  & 6.6  & 4.4 & 3.3 & 2.7 \\
0.86 au & 14.5 & 9.7  & 7.3  & 5.8  & 3.9 & 2.9 & 2.3 \\
0.88 au & 12.5 & 8.3  & 6.2  & 5.0  & 3.3 & 2.5 & 2.0 \\
0.90 au & 10.4 & 6.9  & 5.2  & 4.2  & 2.8 & 2.1 & 1.7 \\
0.92 au & 8.3  & 5.5  & 4.2  & 3.3  & 2.2 & 1.7 & 1.3 \\
0.94 au & 6.2  & 4.2  & 3.1  & 2.5  & 1.7 & 1.2 & 1.0 \\
\end{tabular}%
}
\caption{Forecast lead time in hours as a function of minimum heliocentric distance $r_{min}$ of a DRO and solar wind speed, assuming radial propagation with constant speed and a DRO spacecraft positioned at the Sun--Earth line. This corresponds to the data shown in Figure~4. }
\label{tab:leadtime}
\end{table}


\subsection{Number of spacecraft}

\begin{figure*}
\centering
{\includegraphics[width=\textwidth]{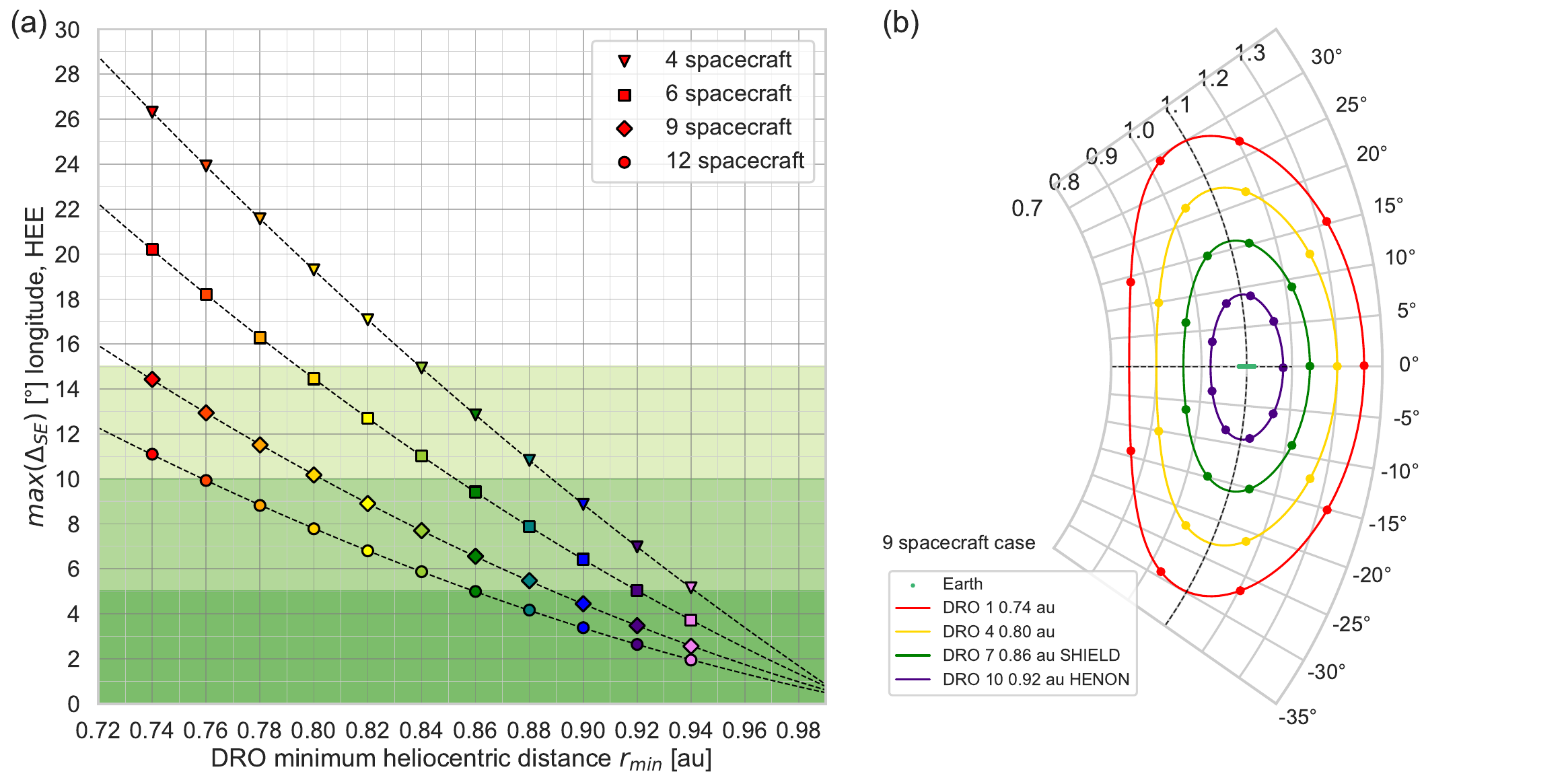}}
\caption{Gap analysis. (a) Maximum difference $\text{max}(\Delta_{SE}$) of a DRO spacecraft in HEE longitude to the Sun--Earth line, for 11 different DROs covering the range 0.74--0.94 au and a given spacecraft configuration or 4, 6, 9, or 12 spacecraft. Different domains with 5, 10 or 15 degrees separation to the Sun-Earth line are marked in different shades of green. (b) Four different DROs as examples (0.74, 0.80, 0.86, 0.92 au) with a 9 spacecraft configuration, at the time of maximum separation to the Sun-Earth line.
} \label{fig:gap}
\end{figure*}

\begin{figure*}
\centering
{\includegraphics[width=\textwidth]{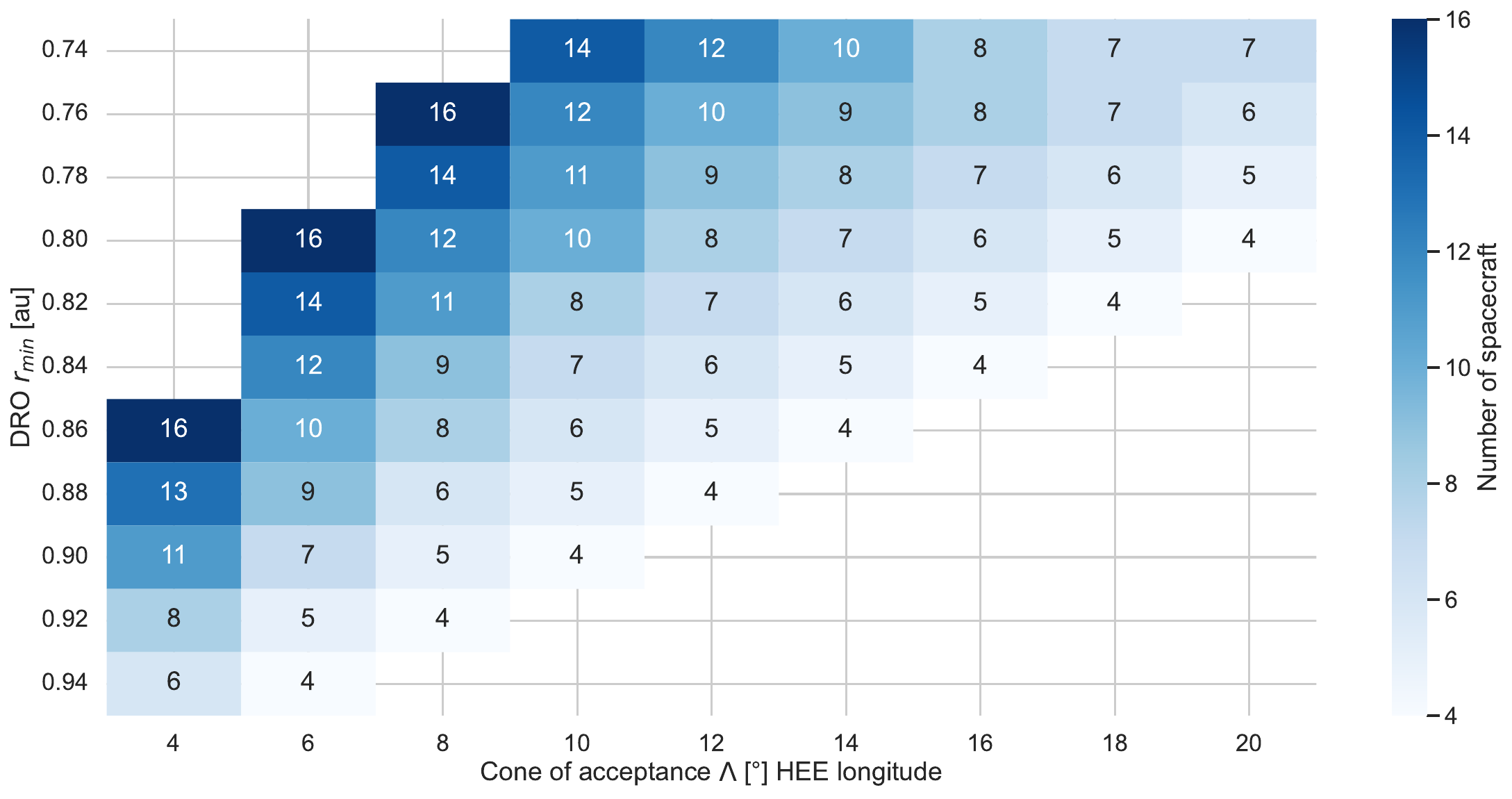}}
\caption{Number of spacecraft needed. This heatmap shows the minimum number of spacecraft in order to fulfill the requirement that always one spacecraft is at a smaller heliospheric longitude $\theta < \Lambda$ to Earth (either east or west), in a grid given by the cone of acceptance $\Lambda$ in steps of 2 degrees, and for 11 DROs with an $r_{min}$ between 0.74 and 0.94 au, in 0.02 au steps. We assume a minimum of 4 spacecraft.
} \label{fig:gap_heatmap}
\end{figure*}

A major question for any given DRO configuration concerns how many spacecraft are needed so that the ``mesh" formed by the spacecraft on the DRO is dense enough to not (1) let potentially strongly geoeffective events slip through undetected, and (2) to ensure that there is enough coherence between the in situ measurements at the sub-L1 point and then later at the Sun--Earth L1 location. This depends on the 3D magnetic structure and shape of CMEs. The most important characteristic here concerns the intervals between spacecraft at different DROs when they pass the Sun--Earth (SE) line, and their maximum distance in longitude from the SE line. 

Figure~\ref{fig:gap}a gives different scenarios for spacecraft fleet configurations with 4, 6, 9 and 12 spacecraft. In Figure~\ref{fig:gap}b the case for 9 spacecraft is demonstrated, at the point in time when the two spacecraft closest to the Sun reach the maximum separation $max(\Delta_{SE})$ to the Sun--Earth line. Table~\ref{tab:gap} presents the main results shown in Figure~\ref{fig:gap}a in numerical form. 

For example, if the cone of acceptance is chosen as $\Lambda=10$°, a 12-spacecraft constellation could be placed at $r_{min}$=0.76 au, one consisting of 9 spacecraft at 0.805 au, one with 6 spacecraft at 0.855 au and one with 4 spacecraft at 0.9 au. For the lead time, the 6 spacecraft constellation at 0.86 au would give 5.8 hours for CMEs with speeds of 1000~\kms, compared to about 25 minutes for L1 based forecasts. 


Figure~\ref{fig:gap_heatmap} shows a heatmap plot for the number of spacecraft needed to fulfill the requirement that always one spacecraft is positioned within a longitude of $\pm \Lambda$ from the Sun--Earth line, with $\Lambda$ being the longitude defined as the ``cone of acceptance", to be set from results on CME coherency and thus to be determined during the mission planning phase.

The main result here is that when raising the number of spacecraft, we reach a regime of diminished returns. The difference in the gap to the Sun--Earth line is getting less pronounced when making the step e.g.\ from 9 to 12 spacecraft, than from 6 to 9 spacecraft. For example, as shown above, for the requirement that no spacecraft should ever be further away than 10° from the Sun--Earth line, 6 spacecraft would be sufficient for a SHIELD like orbit at 0.86 au. However, it is still to be determined what such an acceptable maximum separation to the Sun--Earth line exactly would be, and for reasons of redundancy it is clearly desirably to plan with a higher number of spacecraft than the number that exactly fulfills the requirement.

We have chosen a minimum number of four spacecraft. This constraint is based on the following reasoning: for one and two spacecraft on any DRO, no continuous monitoring is possible with sufficient upstream distances. This is acceptable for a pathfinder mission such as HENON, but is not feasible for a 24/7 operational space weather mission. For three spacecraft, there are times when one spacecraft is behind the Earth and the other two are situated also not far upstream of Earth, inhibiting continuous monitoring by taking advantage of the upstream monitoring concept. We thus use a minimum acceptable number of four spacecraft.

From Figure~\ref{fig:gap_heatmap} it emerges, consistent with everything discussed previously in this study, that for a given $\Lambda$, DROs that are closer to the Sun become wider. To counteract the fact that the gaps between spacecraft become wider too, one needs more spacecraft to fulfill the same $\Lambda$ requirement. Similarly, the smaller $\Lambda$ becomes for the same $r_{min}$, the number of spacecraft needs to increase for denser coverage. It emerges as a rule of thumb that for dividing $\Lambda$ by 2, the number of spacecraft roughly doubles. For an example at 0.86 au, with $\Lambda$=12°, 5 spacecraft are sufficient, whereas with $\Lambda$=6°, 10 spacecraft are needed. Decreasing $r_{min}$ by 0.02 au roughly often leads to exactly one more spacecraft, at least for values of $\Lambda >10$°. For smaller values $\Lambda <10$° and smaller heliocentric distances $r_{min}$ of the DRO, there is a stronger increase in the number of spacecraft needed. In general, we derive from Figure~\ref{fig:gap_heatmap} that on the lower right of the figure the number of spacecraft needed does not depend much on small changes in the mission setup, but this changes more quickly towards the upper left of the figure---this is the domain where adding more spacecraft leads to diminished returns.

 
\begin{table}[h]
\centering
\begin{tabular}{c|cccc}
\toprule
 & 4 s/c & 6 s/c & 9 s/c & 12 s/c \\
\hline
0.74 au & 26.3 & 20.2 & 14.4 & 11.1 \\
0.76 au & 23.9 & 18.2 & 12.9 & 9.9  \\
0.78 au & 21.6 & 16.3 & 11.5 & 8.8  \\
0.80 au & 19.3 & 14.4 & 10.2 & 7.8  \\
0.82 au & 17.1 & 12.7 & 8.9  & 6.8  \\
0.84 au & 14.9 & 11.0 & 7.7  & 5.9  \\
0.86 au & 12.8 & 9.4  & 6.6  & 5.0  \\
0.88 au & 10.8 & 7.9  & 5.5  & 4.2  \\
0.90 au & 8.9 & 6.4  & 4.4  & 3.4  \\
0.92 au & 7.0  & 5.0  & 3.5  & 2.6  \\
0.94 au & 5.2  & 3.7  & 2.6  & 1.9  \\
\end{tabular}
\caption{The maximum separation $max(\Delta_{SE})$ (in degrees) to the Sun--Earth line, of a spacecraft as part of a fleet, for 11 different DROs with a given $r_{min}$ in each row and four different numbers of equidistant distributed spacecraft along a DRO in the columns. 
}
\label{tab:gap}
\end{table}




\section{Discussion} \label{sec:discussion}

We now move on to discuss novel definitions concerning the spatial domains involved in upstream monitoring, and see where spacecraft have already made ICME observations with respect to the Sun--Earth line. 

\subsection{Sub-L1 and far upstream monitoring}

\begin{figure*}
\centering
{\includegraphics[width=16 cm]{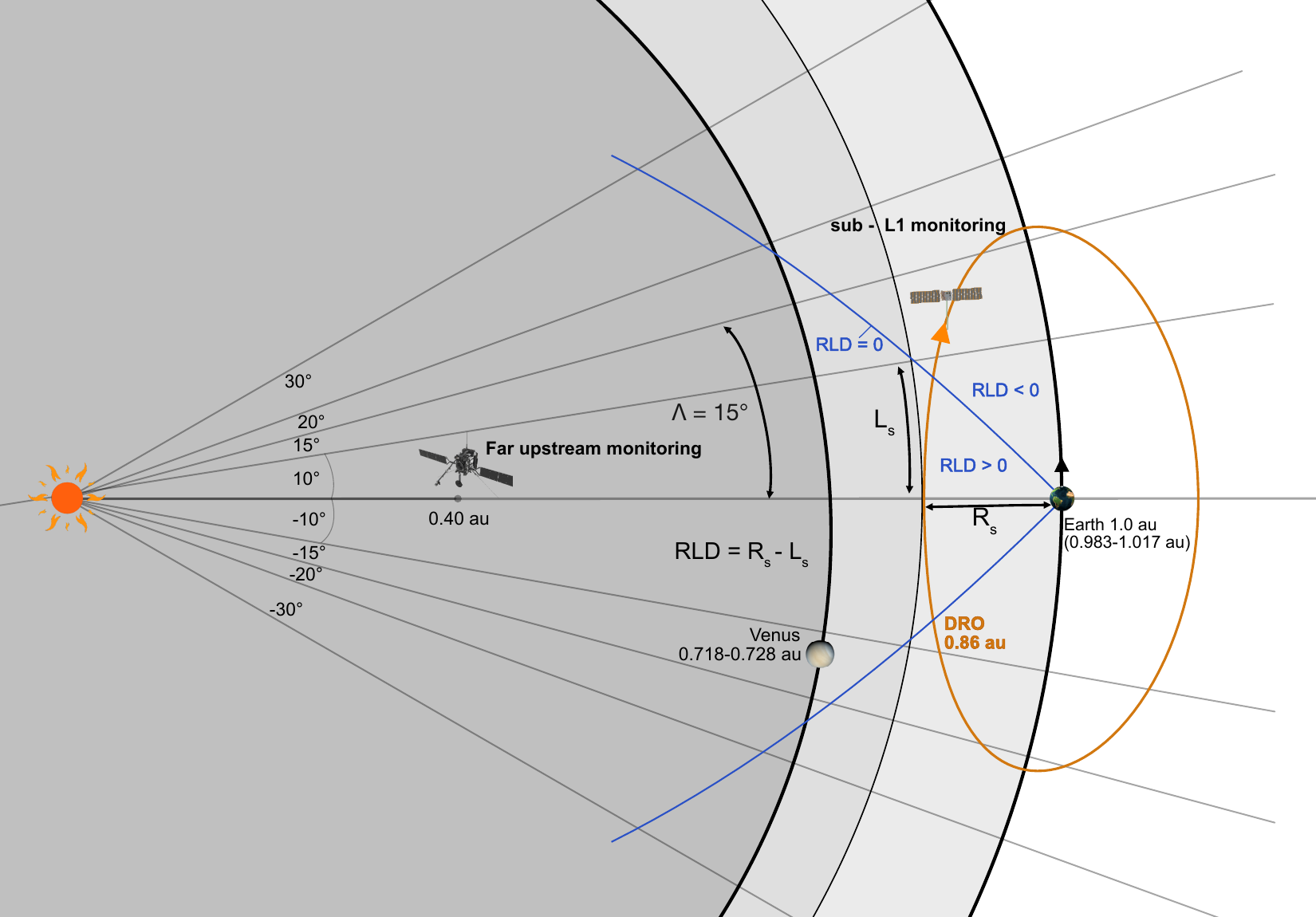}}
\caption{Illustration of our definition of far upstream monitoring and sub-L1 monitoring, with Venus acting as the discriminator between the two regimes. A distant retrograde orbit for the SHIELD concept, with a minimum distance of 0.86 au, is given as an orange solid line. Another concept is introduced as the blue solid lines, which shows the points for which the distance to the Sun--Earth line ($L_s$) is similar as the distance to 1 au ($R_s$). We call this the RLD $=0$ line, meaning the radial to longitude spatial scale difference is 0. The line discriminates the domains for which radial evolution of an ICME matters more (RLD $> 0$) and the domain for which the longitudinal structure of an ICME is of higher importance (RLD $< 0$). A ``cone of acceptance,'' with an assumed width of $\Lambda$ of 15°, is indicated too.} 
\label{fig:dro_sketch}
\end{figure*}
Figure~\ref{fig:dro_sketch} illustrates an overview of the different domains that are involved in upstream monitoring and introduces the distinction between ``sub-L1'' and ``far upstream'' monitoring that we propose here. We define ``sub-L1" monitoring as operating within the heliocentric distance range accessible to DROs, with minimum heliocentric distances of $ > 0.73$~au, just beyond the orbit of Venus (seen from the Sun). In contrast, we use ``far upstream'' monitoring for distances $< 0.71$~au, which is a distance that is, in the long term, not accessible to DROs because of the presence of Venus. Another discriminator, which is roughly consistent with this definition, arises from the fact that an ICME passes a spacecraft fully at around $< 0.75$ au before impacting L1 \citep[e.g.][]{Davies_2021_solo, regnault2023}. For sub-L1 locations, we expect that often an ICME will start to impact L1 while it is still being sampled by the sub-L1 spacecraft, whereas for far upstream distances, the full ICME will be observed and its complete in situ data are potentially available if they have already been transmitted \citep{laker2024,davies2026_solo}.


\subsection{Radial and longitudinal domains}

We move on to introduce another novel concept concerning two different domains, now by looking at the differences in spatial scales that are relevant for a sub-L1 monitor. There are two main spatial scales of importance, concerning its heliospheric position with respect to (1) the longitude difference to the Sun--Earth line and (2) the radial distance to 1 au. In principle, latitudinal differences can also play a role, given that the ecliptic is tilted to the solar equator, but we omit them for simplicity here. These scales determine largely what types of physics are relevant in order to map the sub-L1 in situ data in an optimal way to the L1 point, to obtain the best possible correlation with the in situ data at L1 and to generate accurate forecasts. 
Looking again at Figure~\ref{fig:dro_sketch}, for each point within the domain $[0.3,1]$~au and $[-30,30]$° heliospheric longitude we define a radial--longitudinal difference RLD (in units of au), with $R$ being the heliocentric distance of a given spacecraft position and $\theta$ its heliospheric longitude, as:
\begin{align}
R_s [\text{au}] & = 1 - R [\text{au}] \\
L_s [\text{au}] & = 2 \pi R [\text{au}] \times \text{abs}(\theta [\deg])/360 \\ 
\mathrm{RLD} & = R_s - L_s
\end{align}
with $R_s$ being the relevant radial scale size, and $L_s$ the longitudinal scale size.
One can see that far upstream monitoring exclusively happens in the RLD $>0$ radial domain. For example, in \cite{laker2024} two CMEs were observed in March 2022 at 0.48 (0.44) au, with a difference of 2.3° (9.6°) longitude to the Sun--Earth line, resulting in a RLD of 0.501 (0.486) au. \cite{davies2026_solo} discuss two ICME events in March 2024 at 0.45 (0.39) au and -4.1°(9.6°) with RLDs of 0.518 (0.545) au. Given the radial scale size for these events observed by Solar Orbiter dominates with a distance of about 0.5 au over the longitudinal one, it is clear that in these and similar cases radial CME evolution effects will be of high importance, and the mapping of the in situ data can focus on the radial evolution, including power laws for the magnetic field, kinematics modeling, and the expansion of the ICME. 

For sub-L1 monitoring, the situation is more complex: for a potential SHIELD orbit with $r_{min}=0.86$ au, as shown in Figure~\ref{fig:dro_sketch}, near the Sun--Earth line it is clear that the radial evolution again dominates, however, at around 8° heliospheric longitude (or about 42° opening angle as seen from Earth) this orbit crosses the RLD $=0$ line, and until its maximum extension in longitude $\theta$, the longitude scale $L_s$ dominates. Based on STEREO-A’s recent crossing of the Sun--Earth line, \cite{weiler2026} conducted a statistical analysis of CME events measured by STEREO-A at a sub-L1 position at ${\sim}0.96$~au. The trajectory of STEREO-A was not symmetrical with respect to the Sun--Earth line and therefore did not exactly mimic a DRO. However, as for the RLD parameter, since STEREO-A was relatively close to Earth, it never exceeded $0.06$~au for the time investigated. Furthermore, when STEREO-A was at a heliospheric longitude of $15^\circ$ and radially 0.06~au upstream of Earth, the resulting RLD parameter is $-0.19$~au, indicating that radial evolution effects are mostly less important than longitudinal variations for the investigated STEREO-A crossing of the Sun--Earth line. The authors of this study also conclude that longitudinal separations, determined by the spacecraft position, dominate the differences in the in situ profiles as measured sub-L1 and at the L1 point rather than global evolution. \cite{weiss2024distorted} model an ICME observation on 2023 April 23 at $-$10.1° and 0.05 au difference to L1, giving an RLD of $-0.12$~au. They successfully fitted a deformable flux rope model simultaneously to the separated in situ observations, with a resulting slightly deformed flux rope, emphasizing that longitudinal structure of ICMEs needs to be included for RLD $<0$ setups.

This means that for a SHIELD like mission at 0.86 au, the physics of both longitudinal structure and radial evolution should be considered in order to get the best results for L1 forecasting. Most of the time, this mission would stay in the RLD $> 0$ domain, but at 8° longitude, which is a reasonable cone of acceptance value, it does cross RLD $= 0$.

\subsection{ICME coverage}

\begin{figure*}
\centering
{\includegraphics[width=15 cm]{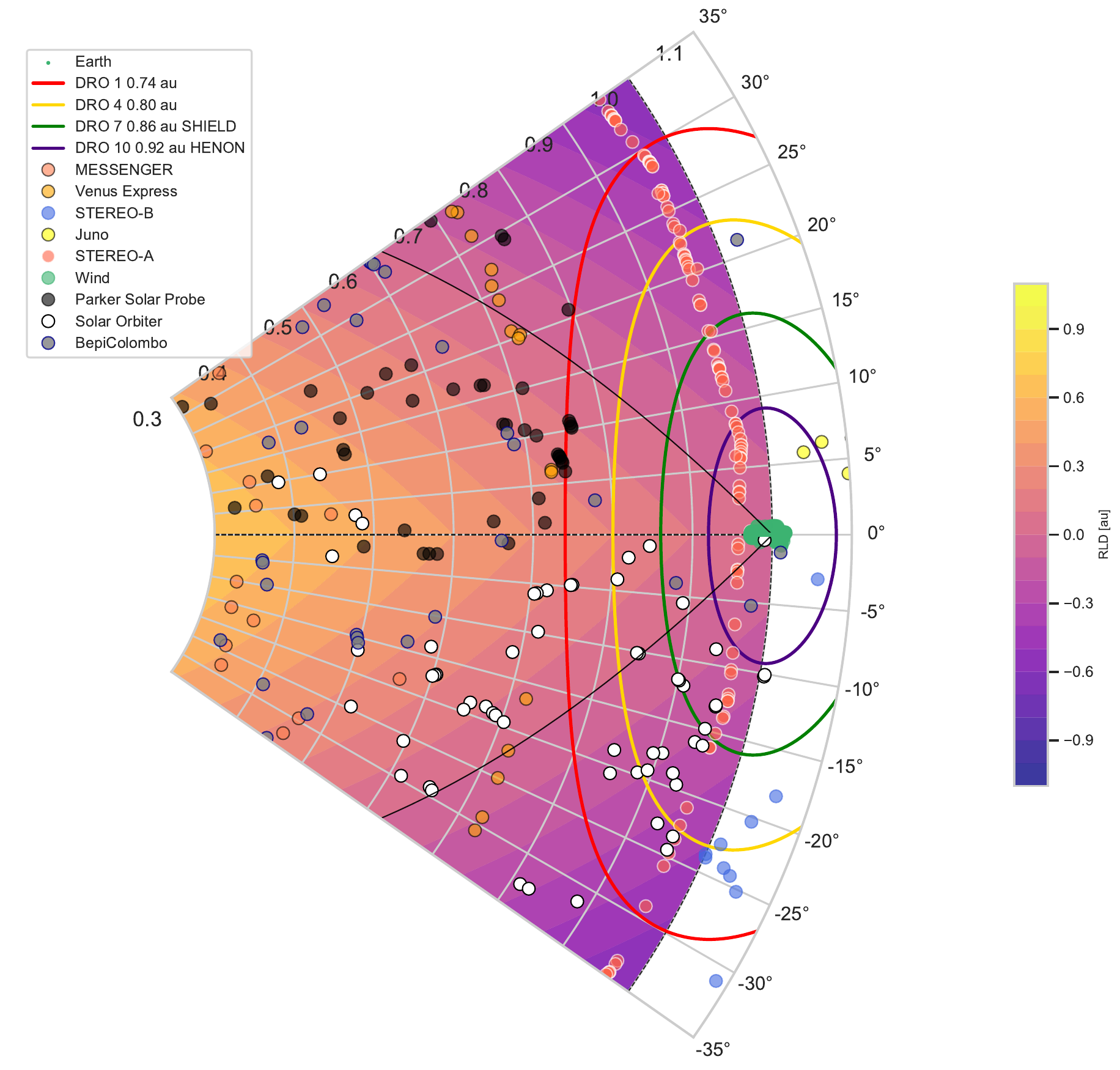}}
\caption{Illustration of in situ observed ICME events with respect to the two radial-longitude-difference (RLD) domains. For RLD $\gg 0$, for which the understanding of the radial evolution dominates, some events have been observed near the Sun--Earth line by MESSENGER (orange dots), Solar Orbiter (white), Parker Solar Probe (black) and BepiColombo (grey). For RLD $\ll 0$, STEREO-A (red) and Solar Orbiter (white) have provided data near 1 au, although with Solar Orbiter so far only on the east side of Earth (longitude $< 0$°). } 
\label{fig:dro_rld}
\end{figure*}
Given the different domains that we introduced, it is of high interest to ask which ICME observations are already available at positions near the Sun--Earth line since 1995, as since then we have a continuous coverage of the L1 solar wind provided by spacecraft such as ACE and Wind, and whether there are still observational gaps. An overview of case studies discussed in the literature has been shown by \cite{palmerio2025monitor}, and here we want to assess the complete set of ICME observations so far available.

Figure \ref{fig:dro_rld} demonstrates the ICME observations so far with respect to the radial and longitudinal domains, which is given as a colormap, using our ICMECAT catalog \citep{moestl2020,moestl2026}. This shows that there is already a substantial historical coverage, which particularly improved due to the launch of Solar Orbiter in 2020 and the fortunate crossing of the Sun--Earth line of STEREO-A between 2022 and 2024 \citep[]{banu2025, weiler2026}. However, little ICME event data is so far available from the branch west of Earth along RLD $=0$, but some events have been observed on the side east of Earth along RLD $=0$. This renders an investigation into east--west asymmetries impossible at the moment. \cite{weiler2026} compared the arrival times of CME shock fronts at STEREO-A with those at L1 while STEREO-A was crossing the Sun--Earth line. The authors observed an east--west asymmetry in lead time, with events observed by STEREO-A east of Earth tending to have longer lead times than those observed west of Earth. This suggests that CMEs and shock fronts are deformed by the interplanetary magnetic field during their propagation. However, since STEREO-A's passage was not symmetric with respect to the Sun--Earth line, this asymmetry could also be an effect of the trajectory itself. In fact, a (somewhat extreme) contrasting example was reported by \cite{rodriguez-garcia2025}, who studied an ICME observed by Solar Orbiter at 0.92~au and approximately 18\dg east of Earth, which nevertheless reached L1 about 12 hours before impacting Solar Orbiter---further highlighting the importance of longitudinal asymmetries in CME propagation. ESA HENON, to be positioned at a $r_{min}\approx 0.92$ au \citep{cicalo2025henon} and on a DRO that is symmetric around the Sun--Earth line by design, may be able to address the origin of this potentially observed asymmetry and to at least partly fill this observational gap for ICME observations on DROs.

In summary, every ICME observation can be assigned to an RLD value, which immediately tells us what the importance of the longitudinal and radial effects are that we need to take into account when we make forecasts. The case RLD $\gg 0$ means it is radial evolution that we need to focus on, whereas RLD $\ll 0$ means that the longitudunal structure is more important, and RLD $\sim 0$ means both effects are equally valid and need to be taken into account for accurately mapping the sub-L1 or far upstream data to the Sun--Earth L1 point.





\section{Conclusions} 
\label{sec:conclusion}

Spacecraft missions for space weather forecasting on DROs have significant potential to improve prediction lead times while maintaining good accuracy, due to the possibility of directly measuring the southward $B_z$ component of CME magnetic fields in situ. Although remote-sensing techniques at radio wavelengths, such as Faraday rotation and gyrosynchrotron emission, can provide estimates of CME magnetic field properties \citep[e.g.][]{liu2007, carley2017, kooi2023}, these methods are not currently available for routine operational forecasting, making in situ measurements in both the ``sub-L1'' and ``far upstream'' regimes defined in this work particularly valuable.

The concept of DROs has proven to be difficult to demonstrate and visualize based on available studies. This motivated us to provide here a simple numerical method, resulting in plots, animations and analyses concerning DROs for CME forecasting, for several different configurations of DROs. Our results have already proven to be highly useful for presentations, and to spread the general understanding of how missions on DRO trajectories work.

We have demonstrated important characteristics of these orbits, and discussed a novel concept called ``cone of acceptance'' which delimits the longitudinal range of the upstream spacecraft useful for forecasts. We have further shown what range of forecast lead times can be expected, and, most importantly, have shown how many spacecraft one needs to fulfill requirements given by the acceptable gaps between spacecraft near the Sun--Earth line.

A major point concerning a space weather mission using DROs is that there are tradeoffs one needs to balance. While it is in principle desirable to go as close to the Sun as possible to extend the forecast lead time, the DROs become wider in longitude, and thus the number of spacecraft needed to make the grid dense enough rises, and so does the mission cost. For higher budgets, one could build more spacecraft and bring them in orbits closer to the Sun, resulting in longer lead times. An optimal DRO mission configuration is thus also determined by how much money we would like to spend on it.
We have also shown that for sub-L1 monitoring to work at useful distances, from around 0.8 to 0.9 au, we need to be aware of both longitudinal effects on CME structure and their radial evolution. Far upstream monitoring ($< 0.71$~au), which has been done for some events with Solar Orbiter \citep{laker2024, davies2026_solo}, can help in guiding us with results concerning the DROs near the sub-L1 lower limit, between around 0.75 to 0.8 au.

There are a few open questions concerning the exact range of the cone of acceptance, in which a spacecraft is in a golden zone for providing in situ data for reliable upstream forecasting results. In particular, research on CME coherence \citep{owens2017} and radial lineup studies with multipoint observations \citep{davies2021_juno, regnault2024} are needed for a better understanding of this issue. Another avenue to be studied are the potential effects of an east--west asymmetry, driven by solar rotation - it might be possible that not the point exactly on the Sun--Earth line correlates best with L1, but slightly offset. Then, ideally before sub-L1 missions are realized, research on the driving of magnetospheric simulations with sub-L1 data, and the derivation of various geomagnetic indices for real time forecasts of geomagnetic effects should be carried out.

In summary, we need to better understand how effective space weather forecasting can be done with sub-L1 monitors, by constraining the trade off between lead time and forecasting accuracy for different distances and longitude differences to Earth.


\section*{Data Availability Statement}
The source code for producing the results, figures and animations for this paper is available at \url{https://github.com/cmoestl/dro_orbits}. An example notebook on how to read the pre-calculated orbits in HEE coordinates is available in the github repository (example.ipynb). The source code to produce the results in this paper is released as version 0.1 at \url{https://github.com/cmoestl/dro_orbits/releases/tag/v0.1}. Animations of different numbers of spacecraft on different DROs, similar to Figure 2, can be found at \url{https://doi.org/10.6084/m9.figshare.32956286}.  The ICMECAT catalog is available at \url{https://helioforecast.space/icmecat} and \url{doi: 10.6084/m9.figshare.6356420}. The version v24 on figshare from 2025 October 15 has been used in this study.

\begin{acknowledgments}
This work is supported by ERC grant (HELIO4CAST, 10.3030/101042188). Funded by the European Union. Views and opinions expressed are however those of the author(s) only and do not necessarily reflect those of the European Union or the European Research Council Executive Agency. Neither the European Union nor the granting authority can be held responsible for them. N.L.\ acknowledges support from NASA 80NSSC24K1245. E.P.\ acknowledges support from NSF SHINE (grant No.\ AGS-2301403).
\end{acknowledgments}


\appendix
\section{Derivation of the Equations of motion} \label{app:a}

\begin{figure*}
\centering
{\includegraphics[width=10 cm]{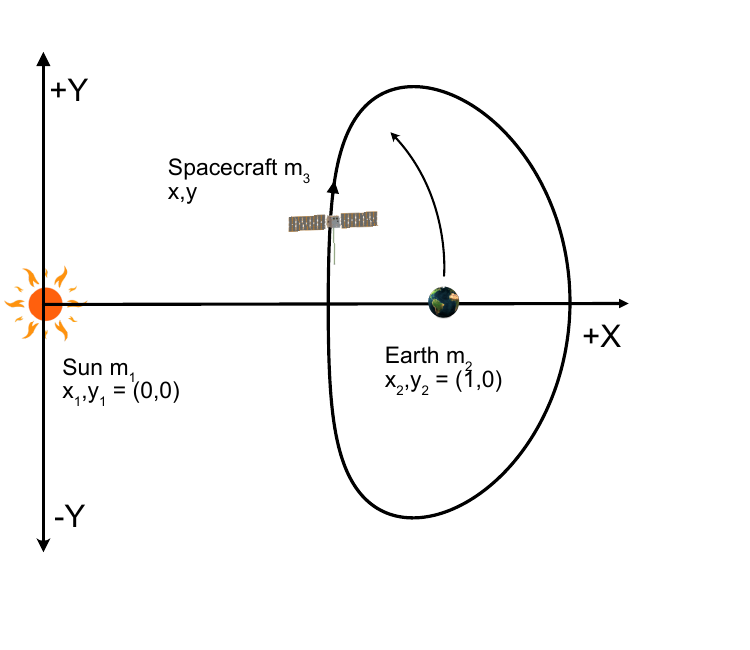}}
\caption{The Heliocentric Earth Ecliptic coordinate system used for the CR3BP calculations. Directions of motion are shown for a view from ecliptic North ($+Z$). } 
\label{fig:appendix_coord}
\end{figure*}

We use the formulation of the circular constrained three body problem (CR3BP) in \cite{Frnka2010}. The coordinate system is illustrated in Figure \ref{fig:appendix_coord}.

The position of the spacecraft given in a cartesian coordinate system as \(x=x(t)\), \(y=y(t)\), with the Sun at $(x_1,y_1)=(0,0)$ and the Earth stationary at $(x_2,y_2)=(1,0)$, which means we now move to a frame in which the Earth is at a fixed position. The angular velocity of the Earth's rotation around the Sun is given as $\omega= 1.990987 \times 10^{-7}\, \text{rad~s}^{-1}$. This is derived from Keplers third law for a circular orbit $\omega = \sqrt{{Gm_{1}}/{r^3}}$, where and $m_1=1.989 \times 10^{30} \, \text{kg}$ the mass of the Sun, and $r$ is the distance of 1~au in units of km. The mass $m_2$ is the Earth, and $m_3$ the spacecraft. An important assumption of the CR3BP is that the heliocentric distance of the Earth is always at 1 au, which is quite well fulfilled, given the Earth's periohelion and aphelion vary between 0.983 and 1.017 au. The coordinate transformation for the fixed frame including the Earth's rotation around the Sun is:

$$\begin{pmatrix}
x'\\
y'
\end{pmatrix}
= 
\begin{pmatrix}
\cos(\omega t) & -\sin(\omega t)\\
\sin(\omega t) & \cos(\omega t)
\end{pmatrix}
\begin{pmatrix}
x\\
y
\end{pmatrix}$$

$$\begin{cases}
  x' &= x\cos(\omega t) - y\sin(\omega t)\\
  y' &= y\cos(\omega t) + x\sin(\omega t)
\end{cases}$$

$$\begin{cases}
  \dot{x}' &= \dot{x}\cos(\omega t) - \omega x\sin(\omega t) - \dot{y}\sin(\omega t) - \omega y\cos(\omega t)\\
  \dot{y}' &= \dot{y}\cos(\omega t) - \omega y\sin(\omega t) + \dot{x}\sin(\omega t) + \omega x\cos(\omega t)
\end{cases}$$

$$\begin{cases}
  \dot{x}' &= \cos(\omega t)(\dot{x} - \omega y) 
  - \sin(\omega t)(\dot{y} + \omega x) \\
  \dot{y}' &= \cos(\omega t)(\dot{y} + \omega x) + \sin(\omega t)(\dot{x} - \omega y)
\end{cases}$$

$$
\begin{cases}
\begin{aligned}
&(\dot{x}')^2 = \cos^2(\omega t)(\dot{x} - \omega y)^2 
  + \sin^2(\omega t)(\dot{y} + \omega x)^2 -2\sin(\omega t)\cos(\omega t)(\dot{x} - \omega y)(\dot{y} + \omega x)\\
& (\dot{y}')^2 = \cos^2(\omega t)(\dot{y} + \omega x)^2 + \sin^2(\omega t)(\dot{x} - \omega y)^2 + 2\sin(\omega t)\cos(\omega t)(\dot{x} - \omega y)(\dot{y} + \omega x)    
\end{aligned}
\end{cases}
$$

The Lagrangian is 

\begin{equation}
    L=T-V
\end{equation}

\begin{align*}
    T &=\dfrac{1}{2}m_3((\dot{x}')^2+(\dot{y}')^2) \\
    &= \dfrac{1}{2}m_3((\dot{x} - \omega y)^2+(\dot{y} + \omega x)^2)\\
    &= \dfrac{1}{2}m_3(\dot{x}^2+\dot{y}^2
    -2y\omega \dot{x} +2x\omega \dot{y}
    +\omega^2(x^2+y^2))
\end{align*}

With \(V\) being
\begin{equation}
    V = -\dfrac{Gm_1m_3}{p_1}-\dfrac{Gm_2m_3}{p_2},
\end{equation}where \(p_1=\sqrt{(x-x_1)^2+(y-y_1)^2}\) and \(p_2=\sqrt{(x-x_2)^2+(y-y_2)^2}\)
the Lagrangian is

\begin{equation*}
    L = \dfrac{1}{2}m_3(\dot{x}^2+\dot{y}^2
    -2y\omega \dot{x} +2x\omega \dot{y}
    +\omega^2(x^2+y^2)) + \dfrac{Gm_1m_3}{p_1} + \dfrac{Gm_2m_3}{p_2} 
\end{equation*}  
\\

\noindent Considering that 
\begin{equation*}
    \dfrac{\partial p_1}{\partial x} = \dfrac{x-x_1}{p_1} \quad \quad \dfrac{\partial p_2}{\partial x} = \dfrac{x-x_2}{p_2}
\end{equation*}and
\begin{equation*}
    \dfrac{\partial p_1}{\partial y} = \dfrac{y-y_1}{p_1} \quad \quad \dfrac{\partial p_2}{\partial y} = \dfrac{y-y_2}{p_2}
\end{equation*}

\noindent Then
\begin{equation*}
    \dfrac{\partial L}{\partial \dot{x}} = \dfrac{1}{2}m_3(2\dot{x}-2y\omega) = m_3(\dot{x}-\omega y) \quad \Rightarrow \quad \dfrac{d}{dt}\biggl(\dfrac{\partial L}{\partial \dot{x}}\biggl) = m_3(\ddot{x}-\omega \dot{y})
\end{equation*}
\begin{equation*}    
    \dfrac{\partial L}{\partial \dot{y}} = \dfrac{1}{2}m_3(2\dot{y}+2x\omega) = m_3(\dot{y}+\omega x) \quad \Rightarrow \quad \dfrac{d}{dt}\biggl(\dfrac{\partial L}{\partial \dot{y}}\biggl) = m_3(\ddot{y}+\omega \dot{x})
\end{equation*}
And considering that
\begin{align*}
    \dfrac{\partial L}{\partial x} &= \dfrac{1}{2}m_3(2\omega \dot{y} + 2 \omega^2 x) - \dfrac{Gm_1m_3}{p_1^2}\dfrac{\partial p_1}{\partial x} - \dfrac{Gm_2m_3}{p_2^2}\dfrac{\partial p_2}{\partial x} \\
    &= \dfrac{1}{2}m_3(2\omega \dot{y} + 2 \omega^2 x) - \dfrac{Gm_1m_3}{p_1^3}(x-x_1) - \dfrac{Gm_2m_3}{p_2^3}(x-x_2) \\
    &= m_3\omega \dot{y} + m_3\omega^2 x - \dfrac{Gm_1m_3}{p_1^3}(x-x_1) - \dfrac{Gm_2m_3}{p_2^3}(x-x_2)
\end{align*} 
\begin{align*}
    \dfrac{\partial L}{\partial y} &= \dfrac{1}{2}m_3(-2\omega \dot{x} + 2 \omega^2 y) - \dfrac{Gm_1m_3}{p_1^2}\dfrac{\partial p_1}{\partial y} - \dfrac{Gm_2m_3}{p_2^2}\dfrac{\partial p_2}{\partial y} \\
    &= \dfrac{1}{2}m_3(-2\omega \dot{x} + 2 \omega^2 y) - \dfrac{Gm_1m_3}{p_1^3}(y-y_1) - \dfrac{Gm_2m_3}{p_2^3}(y-y_2) \\
    &= -m_3\omega \dot{x} +  m_3\omega^2 y - \dfrac{Gm_1m_3}{p_1^3}(y-y_1) - \dfrac{Gm_2m_3}{p_2^3}(y-y_2)
\end{align*}

\noindent Then the Lagrange's equations
$$
\begin{cases}
    \dfrac{d}{dt}\biggl(\dfrac{\partial L}{\partial \dot{x}}\biggl) - \dfrac{\partial L}{\partial x} = 0 \\
    \dfrac{d}{dt}\biggl(\dfrac{\partial L}{\partial \dot{y}}\biggl) - \dfrac{\partial L}{\partial y} = 0
\end{cases}
$$
can be easily written as
$$
\begin{cases}
    m_3(\ddot{x}-\omega \dot{y}) = m_3\omega \dot{y} + m_3 \omega^2 x - \dfrac{Gm_1m_3}{p_1^3}(x-x_1) - \dfrac{Gm_2m_3}{p_2^3}(x-x_2) \\
    m_3(\ddot{y}+\omega \dot{x}) = m_3(-\omega \dot{x} + \omega^2 y) - \dfrac{Gm_1m_3}{p_1^3}(y-y_1) - \dfrac{Gm_2m_3}{p_2^3}(y-y_2)
\end{cases}
$$

$$
\begin{cases}
    m_3\ddot{x} = 2 m_3\omega \dot{y} + m_3\omega^2 x - \dfrac{Gm_1m_3}{p_1^3}(x-x_1) - \dfrac{Gm_2m_3}{p_2^3}(x-x_2) \\
    m_3\ddot{y} = -2m_3\omega \dot{x} + m_3\omega^2 y - \dfrac{Gm_1m_3}{p_1^3}(y-y_1) - \dfrac{Gm_2m_3}{p_2^3}(y-y_2)
\end{cases}
$$

In the last two equations $m_3$ can be cancelled out, and we are left with the final form of the equations of motion in the CR3BP framework:

\begin{equation}
\label{eq:cr3bp}
\begin{cases}
    \ddot{x} = 2 \omega \dot{y} + \omega^2 x - \dfrac{Gm_1}{p_1^3}(x-x_1) - \dfrac{Gm_2}{p_2^3}(x-x_2) \\
    \ddot{y} = -2\omega \dot{x} + \omega^2 y - \dfrac{Gm_1}{p_1^3}(y-y_1) - \dfrac{Gm_2}{p_2^3}(y-y_2).
\end{cases}
\end{equation}

\bibliography{chrisbib}{}

@ARTICLE{horne2026_maystorm_uk,
       author = {{Horne}, R.~B. and {Angling}, M.~J. and {Attrill}, G.~D.~R. and {Beggan}, C. and {Bisi}, M.~M. and {Cannon}, P.~S. and {Clarke}, E. and {Dyer}, C. and {Eastwood}, J.~P. and {Elvidge}, S. and et al.},
        title = "{The May 2024 geomagnetic storm: UK experience and perspective}",
      journal = {Royal Society Open Science},
         year = 2026,
        month = apr,
       volume = {13},
       number = {4},
          eid = {251943},
        pages = {251943},
          doi = {10.1098/rsos.251943},
       adsurl = {https://ui.adsabs.harvard.edu/abs/2026RSOS...1351943H}
}

@ARTICLE{temmer2021_review,
       author = {{Temmer}, Manuela},
        title = "{Space weather: the solar perspective: An update to Schwenn (2006)}",
      journal = {Living Reviews in Solar Physics},
         year = 2021,
        month = dec,
       volume = {18},
       number = {1},
          eid = {4},
        pages = {4},
          doi = {10.1007/s41116-021-00030-3},
archivePrefix = {arXiv},
       eprint = {2104.04261},
 primaryClass = {astro-ph.SR},
       adsurl = {https://ui.adsabs.harvard.edu/abs/2021LRSP...18....4T}
}

@ARTICLE{palmerio2025monitor,
       author = {{Palmerio}, Erika},
        title = "{Monitoring the Solar Wind Before It Reaches L1}",
      journal = {Space Weather},
         year = 2025,
        month = nov,
       volume = {23},
       number = {11},
          eid = {e2025SW004452},
        pages = {e2025SW004452},
          doi = {10.1029/2025SW004452},
archivePrefix = {arXiv},
       eprint = {2511.08463},
 primaryClass = {physics.space-ph},
       adsurl = {https://ui.adsabs.harvard.edu/abs/2025SpWea..2304452P}
}

@techreport{Frnka2010,
  author      = {Frnka, Richard},
  title       = {The Circular Restricted Three-Body Problem},
  institution = {Northern Arizona University},
  year        = {2010},
  type        = {REU Report},
  note        = {Research Experience for Undergraduates, advised by James Swift},
  url         = {https://jan.ucc.nau.edu/~ns46/student/2010/Frnka_2010.pdf}
}

@ARTICLE{moestl2026,
       author = {{M{\"o}stl}, Christian and {Davies}, Emma E. and {Weiler}, Eva and {R{\"u}disser}, Hannah T. and {Amerstorfer}, Ute V. and {Weiss}, Andreas J. and {Reiss}, Martin A. and {Majumdar}, Satabdwa and {Horbury}, Timothy S. and {Bale}, Stuart D. and {Heyner}, Daniel},
        title = "{On the Magnetic Field Evolution of Interplanetary Coronal Mass Ejections from 0.07 to 5.4 au}",
      journal = {\apj},
         year = 2026,
        month = apr,
       volume = {1001},
       number = {1},
          eid = {70},
        pages = {70},
          doi = {10.3847/1538-4357/ae50fe},
archivePrefix = {arXiv},
       eprint = {2512.04730},
 primaryClass = {astro-ph.SR},
       adsurl = {https://ui.adsabs.harvard.edu/abs/2026ApJ..1001...70M}
}

@article{weiler2026,
author = {Weiler, E. and Davies, E. E. and Möstl, C. and Lugaz, N. and Veronig, A. and Bailey, R. L. and Reiss, M. A.},
title = {Investigating Potential Benefits of Future Sub-L1 Missions With STEREO-A},
journal = {Space Weather},
volume = {24},
number = {6},
pages = {e2026SW005033},
doi = {https://doi.org/10.1029/2026SW005033},
url = {https://agupubs.onlinelibrary.wiley.com/doi/abs/10.1029/2026SW005033},
eprint = {https://agupubs.onlinelibrary.wiley.com/doi/pdf/10.1029/2026SW005033},
note = {e2026SW005033 2026SW005033},
year = {2026}
}

@ARTICLE{prete2026,
       author = {{Prete}, G. and {Niemela}, A. and {Poedts}, S. and {Zimbardo}, G. and {Cical{\`o}}, S. and {Marcucci}, M.~F. and {Laurenza}, M. and {Stumpo}, M. and {Landi}, S. and {Sangalli}, M. and et al.},
        title = "{Assessing VBz variations during CME propagation: a preparatory study for the HENON mission using EUHFORIA}",
      journal = {arXiv e-prints},
         year = 2026,
        month = mar,
          eid = {arXiv:2603.27310},
        pages = {arXiv:2603.27310},
          doi = {10.48550/arXiv.2603.27310},
archivePrefix = {arXiv},
       eprint = {2603.27310},
 primaryClass = {astro-ph.SR},
       adsurl = {https://ui.adsabs.harvard.edu/abs/2026arXiv260327310P}
}

@ARTICLE{eastwood_2015_sunjammer,
       author = {{Eastwood}, J.~P. and {Kataria}, D.~O. and {McInnes}, C.~R. and {Barnes}, N.~C. and {Mulligan}, P.},
        title = "{Sunjammer}",
      journal = {Weather},
         year = 2015,
        month = jan,
       volume = {70},
       number = {1},
        pages = {27-30},
          doi = {10.1002/wea.2438},
       adsurl = {https://ui.adsabs.harvard.edu/abs/2015Wthr...70...27E}
}

@ARTICLE{ritter2015,
       author = {{Ritter}, Birgit and {Meskers}, Arjan J.~H. and {Miles}, Oscar and {Ru{\ss}wurm}, Michael and {Scully}, Stephen and {Rold{\'a}n}, Andr{\'e}s and {Hartkorn}, Oliver and {J{\"u}stel}, Peter and {R{\'e}ville}, Victor and {Lupu}, Sorina and {Ruffenach}, Alexis},
        title = "{A Space weather information service based upon remote and in-situ measurements of coronal mass ejections heading for Earth. A concept mission consisting of six spacecraft in a heliocentric orbit at 0.72 AU}",
      journal = {Journal of Space Weather and Space Climate},
         year = 2015,
        month = feb,
       volume = {5},
          eid = {A3},
        pages = {A3},
          doi = {10.1051/swsc/2015006},
       adsurl = {https://ui.adsabs.harvard.edu/abs/2015JSWSC...5A...3R}
}

@ARTICLE{perozzi2017,
       author = {{Perozzi}, Ettore and {Ceccaroni}, Marta and {Valsecchi}, Giovanni B. and {Rossi}, Alessandro},
        title = "{Distant retrograde orbits and the asteroid hazard}",
      journal = {European Physical Journal Plus},
         year = 2017,
        month = aug,
       volume = {132},
       number = {8},
          eid = {367},
        pages = {367},
          doi = {10.1140/epjp/i2017-11644-0},
       adsurl = {https://ui.adsabs.harvard.edu/abs/2017EPJP..132..367P}
}

@ARTICLE{davies2026_solo,
       author = {{Davies}, Emma E. and {Weiler}, Eva and {M{\"o}stl}, Christian and {Horbury}, Timothy S. and {O'Brien}, Helen and {Morris}, Jean and {Crabtree}, Alastair},
        title = "{Real-time prediction of geomagnetic storms using Solar Orbiter as a far upstream solar wind monitor}",
      journal = {Space Weather},
         year = 2026,
        month = aug,
       volume = {24},
          eid = {e2026SW005013},
        pages = {e2026SW005013},
          doi = {10.1029/2026SW005013},
archivePrefix = {arXiv},
       eprint = {2508.13892},
 primaryClass = {physics.space-ph},
       adsurl = {https://ui.adsabs.harvard.edu/abs/2025arXiv250813892D}
}

@ARTICLE{rodriguez-garcia2025,
       author = {{Rodr{\'\i}guez-Garc{\'\i}a}, L. and {G{\'o}mez-Herrero}, R. and {Dresing}, N. and {Balmaceda}, L.~A. and {Palmerio}, E. and {Kouloumvakos}, A. and {Jebaraj}, I.~C. and {Espinosa Lara}, F. and {Roco}, M. and {Palmroos}, C. and {Warmuth}, A. and {Nicolaou}, G. and {Mason}, G.~M. and {Guo}, J. and {Laitinen}, T. and {Cernuda}, I. and {Nieves-Chinchilla}, T. and {Fedeli}, A. and {Lee}, C.~O. and {Cohen}, C.~M.~S. and {Owen}, C.~J. and {Ho}, G.~C. and {Malandraki}, O. and {Vainio}, R. and {Rodr{\'\i}guez-Pacheco}, J.},
        title = "{Solar energetic particles injected inside and outside a magnetic cloud: The widespread solar energetic particle event on 2022 January 20}",
      journal = {\aap},
         year = 2025,
        month = feb,
       volume = {694},
          eid = {A64},
        pages = {A64},
          doi = {10.1051/0004-6361/202452158},
archivePrefix = {arXiv},
       eprint = {2409.04564},
 primaryClass = {astro-ph.SR},
       adsurl = {https://ui.adsabs.harvard.edu/abs/2025A&A...694A..64R}
}

@ARTICLE{liu2007,
       author = {{Liu}, Y. and {Manchester}, IV, W.~B. and {Kasper}, J.~C. and {Richardson}, J.~D. and {Belcher}, J.~W.},
        title = "{Determining the Magnetic Field Orientation of Coronal Mass Ejections from Faraday Rotation}",
      journal = {\apj},
         year = 2007,
        month = aug,
       volume = {665},
       number = {2},
        pages = {1439-1447},
          doi = {10.1086/520038},
archivePrefix = {arXiv},
       eprint = {0705.2780},
 primaryClass = {astro-ph},
       adsurl = {https://ui.adsabs.harvard.edu/abs/2007ApJ...665.1439L}
}

@ARTICLE{kooi2023,
       author = {{Kooi}, Jason and {Wexler}, David and {Jensen}, Elizabeth and {Wood}, Brian and {Nieves-Chinchilla}, Teresa and {Manchester}, Ward and {Pevtsov}, Alexei and {Jian}, Lan and {Kenny}, Megan and {Wilson}, Lynn and {Fung}, Shing and {Gopalswamy}, Nat},
        title = "{Probing the Magnetic Field Structure of Coronal Mass Ejections with Faraday Rotation}",
      journal = {BAAS},
         year = 2023,
       volume = {55},
        month = jul,
          eid = {221},
        pages = {221},
          doi = {10.3847/25c2cfeb.0609a250},
       adsurl = {https://ui.adsabs.harvard.edu/abs/2023BAAS...55c.221K}
}

@ARTICLE{carley2017,
       author = {{Carley}, Eoin P. and {Vilmer}, Nicole and {Sim{\~o}es}, Paulo J.~A. and {{\'O} Fearraigh}, Br{\'\i}an},
        title = "{Estimation of a coronal mass ejection magnetic field strength using radio observations of gyrosynchrotron radiation}",
      journal = {\aap},
         year = 2017,
        month = dec,
       volume = {608},
          eid = {A137},
        pages = {A137},
          doi = {10.1051/0004-6361/201731368},
archivePrefix = {arXiv},
       eprint = {1709.05184},
 primaryClass = {astro-ph.SR},
       adsurl = {https://ui.adsabs.harvard.edu/abs/2017A&A...608A.137C}
}

@ARTICLE{henon1969,
       author = {{Henon}, M.},
        title = "{Numerical exploration of the restricted problem, V}",
      journal = {\aap},
         year = 1969,
        month = feb,
       volume = {1},
        pages = {223-238},
       adsurl = {https://ui.adsabs.harvard.edu/abs/1969A&A.....1..223H}
}

@ARTICLE{weiss2024distorted,
       author = {{Weiss}, Andreas J. and {Nieves-Chinchilla}, Teresa and {M{\"o}stl}, Christian},
        title = "{Distorted Magnetic Flux Ropes within Interplanetary Coronal Mass Ejections}",
      journal = {\apj},
         year = 2024,
        month = nov,
       volume = {975},
       number = {2},
          eid = {169},
        pages = {169},
          doi = {10.3847/1538-4357/ad7940},
archivePrefix = {arXiv},
       eprint = {2406.13022},
 primaryClass = {astro-ph.SR},
       adsurl = {https://ui.adsabs.harvard.edu/abs/2024ApJ...975..169W}
}

@ARTICLE{cyr2000diamond,
       author = {{St. Cyr}, O.~C. and {Mesarch}, M.~A. and {Maldonado}, H.~M. and {Folta}, D.~C. and {Harper}, A.~D. and {Davila}, J.~M. and {Fisher}, R.~R.},
        title = "{Space Weather Diamond: a four spacecraft monitoring system}",
      journal = {Journal of Atmospheric and Solar-Terrestrial Physics},
         year = 2000,
        month = sep,
       volume = {62},
       number = {14},
        pages = {1251-1255},
          doi = {10.1016/S1364-6826(00)00069-9},
       adsurl = {https://ui.adsabs.harvard.edu/abs/2000JASTP..62.1251S}
}

@ARTICLE{cicalo2025henon,
       author = {{Cical{\`o}}, Stefano and {Alessi}, Elisa Maria and {Provinciali}, Lorenzo and {Amabili}, Paride and {Saita}, Giorgio and {Calcagno}, Davide and {Marcucci}, Maria Federica and {Laurenza}, Monica and {Zimbardo}, Gaetano and {Landi}, Simone and {Walker}, Roger and {Khan}, Michael},
        title = "{Mission analysis for the HENON CubeSat mission to a large Sun-Earth distant retrograde orbit}",
      journal = {\apss},
         year = 2025,
        month = aug,
       volume = {370},
       number = {8},
          eid = {83},
        pages = {83},
          doi = {10.1007/s10509-025-04473-0},
archivePrefix = {arXiv},
       eprint = {2508.02138},
 primaryClass = {astro-ph.EP},
       adsurl = {https://ui.adsabs.harvard.edu/abs/2025Ap&SS.370...83C}
}

@ARTICLE{weiler2025,
       author = {{Weiler}, E. and {M{\"o}stl}, C. and {Davies}, E.~E. and {Veronig}, A.~M. and {Amerstorfer}, U.~V. and {Amerstorfer}, T. and {Le Lou{\"e}dec}, J. and {Bauer}, M. and {Lugaz}, N. and {Haberle}, V. and {R{\"u}disser}, H.~T. and {Majumdar}, S. and {Reiss}, M.},
        title = "{First Observations of a Geomagnetic Superstorm With a Sub-L1 Monitor}",
      journal = {Space Weather},
         year = 2025,
        month = mar,
       volume = {23},
       number = {3},
        pages = {2024SW004260},
          doi = {10.1029/2024SW004260},
archivePrefix = {arXiv},
       eprint = {2411.12490},
 primaryClass = {physics.space-ph},
       adsurl = {https://ui.adsabs.harvard.edu/abs/2025SpWea..2304260W}
}

@ARTICLE{banu2025,
       author = {{Banu}, Sahanaj A. and {Lugaz}, No{\'e} and {Zhuang}, Bin and {Al-Haddad}, Nada and {Farrugia}, Charles J. and {Galvin}, Antoinette B.},
        title = "{Investigating Coronal Mass Ejections through Multispacecraft Measurements: STEREO-A and L1 in 2022{\textendash}2023}",
      journal = {\apj},
         year = 2025,
        month = mar,
       volume = {982},
       number = {1},
          eid = {47},
        pages = {47},
          doi = {10.3847/1538-4357/adb60c},
       adsurl = {https://ui.adsabs.harvard.edu/abs/2025ApJ...982...47B}
}

@ARTICLE{lugaz2024ssr,
       author = {{Lugaz}, No{\'e} and {Lee}, Christina O. and {Al-Haddad}, Nada and {Lillis}, Robert J. and {Jian}, Lan K. and {Curtis}, David W. and {Galvin}, Antoinette B. and {Whittlesey}, Phyllis L. and {Rahmati}, Ali and {Zesta}, Eftyhia and {Moldwin}, Mark and {Summerlin}, Errol J. and {Larson}, Davin E. and {Courtade}, Sasha and {French}, Richard and {Hunter}, Richard and {Covitti}, Federico and {Cosgrove}, Daniel and {Prall}, J.~D. and {Allen}, Robert C. and {Zhuang}, Bin and {Winslow}, R{\'e}ka M. and {Scolini}, Camilla and {Lynch}, Benjamin J. and {Filwett}, Rachael J. and {Palmerio}, Erika and {Farrugia}, Charles J. and {Smith}, Charles W. and {M{\"o}stl}, Christian and {Weiler}, Eva and {Janvier}, Miho and {Regnault}, Florian and {Livi}, Roberto and {Nieves-Chinchilla}, Teresa},
        title = "{The Need for Near-Earth Multi-Spacecraft Heliospheric Measurements and an Explorer Mission to Investigate Interplanetary Structures and Transients in the Near-Earth Heliosphere}",
      journal = {\ssr},
         year = 2024,
        month = oct,
       volume = {220},
       number = {7},
          eid = {73},
        pages = {73},
          doi = {10.1007/s11214-024-01108-8},
       adsurl = {https://ui.adsabs.harvard.edu/abs/2024SSRv..220...73L}
}

@ARTICLE{lugaz2025need,
       author = {{Lugaz}, No{\'e} and {Al-Haddad}, Nada and {Zhuang}, Bin and {M{\"o}stl}, Christian and {Davies}, Emma E. and {Farrugia}, Charles J. and {Banu}, Sahanaj Aktar and {Weiler}, Eva and {Galvin}, Antoinette B.},
        title = "{The Need for a Sub-L1 Space Weather Research Mission: Current Knowledge Gaps on Coronal Mass Ejections}",
      journal = {Space Weather},
         year = 2025,
        month = feb,
       volume = {23},
       number = {2},
        pages = {2024SW004189},
          doi = {10.1029/2024SW004189},
       adsurl = {https://ui.adsabs.harvard.edu/abs/2025SpWea..2304189L}
}

@ARTICLE{alhaddad2025structure,
       author = {{Al-Haddad}, Nada and {Lugaz}, No{\'e}},
        title = "{The Magnetic Field Structure of Coronal Mass Ejections: A More Realistic Representation}",
      journal = {\ssr},
         year = 2025,
        month = feb,
       volume = {221},
       number = {1},
          eid = {12},
        pages = {12},
          doi = {10.1007/s11214-025-01138-w},
       adsurl = {https://ui.adsabs.harvard.edu/abs/2025SSRv..221...12A}
}

@ARTICLE{palmerio2025event4,
       author = {{Palmerio}, Erika and {Kay}, Christina and {Al-Haddad}, Nada and {Lynch}, Benjamin J. and {Trotta}, Domenico and {Yu}, Wenyuan and {Ledvina}, Vincent E. and {S{\'a}nchez-Cano}, Beatriz and {Riley}, Pete and {Heyner}, Daniel and {Schmid}, Daniel and {Fischer}, David and {Richter}, Ingo and {Auster}, Hans-Ulrich},
        title = "{A coronal mass ejection encountered by four spacecraft within 1 au from the Sun: ensemble modelling of propagation and magnetic structure}",
      journal = {\mnras},
         year = 2025,
        month = jan,
       volume = {536},
       number = {1},
        pages = {203-222},
          doi = {10.1093/mnras/stae2606},
archivePrefix = {arXiv},
       eprint = {2411.12706},
 primaryClass = {astro-ph.SR},
       adsurl = {https://ui.adsabs.harvard.edu/abs/2025MNRAS.536..203P}
}

@ARTICLE{regnault2024,
       author = {{Regnault}, F. and {Al-Haddad}, N. and {Lugaz}, N. and {Farrugia}, C.~J. and {Yu}, W. and {Zhuang}, B. and {Davies}, E.~E.},
        title = "{Discrepancies in the Properties of a Coronal Mass Ejection on Scales of 0.03 au as Revealed by Simultaneous Measurements at Solar Orbiter and Wind: The 2021 November 3{\textendash}5 Event}",
      journal = {\apj},
         year = 2024,
        month = feb,
       volume = {962},
       number = {2},
          eid = {190},
        pages = {190},
          doi = {10.3847/1538-4357/ad1883},
archivePrefix = {arXiv},
       eprint = {2311.14046},
 primaryClass = {astro-ph.SR},
       adsurl = {https://ui.adsabs.harvard.edu/abs/2024ApJ...962..190R}
}

@ARTICLE{lugaz2024_width,
       author = {{Lugaz}, No{\'e} and {Zhuang}, Bin and {Scolini}, Camilla and {Al-Haddad}, Nada and {Farrugia}, Charles J. and {Winslow}, R{\'e}ka M. and {Regnault}, Florian and {M{\"o}stl}, Christian and {Davies}, Emma E. and {Galvin}, Antoinette B.},
        title = "{The Width of Magnetic Ejecta Measured near 1 au: Lessons from STEREO-A Measurements in 2021{\textendash}2022}",
      journal = {\apj},
         year = 2024,
        month = feb,
       volume = {962},
       number = {2},
          eid = {193},
        pages = {193},
          doi = {10.3847/1538-4357/ad17b9},
archivePrefix = {arXiv},
       eprint = {2312.03942},
 primaryClass = {astro-ph.SR},
       adsurl = {https://ui.adsabs.harvard.edu/abs/2024ApJ...962..193L}
}

@ARTICLE{davies2021_juno,
       author = {{Davies}, Emma E. and {Forsyth}, Robert J. and {Winslow}, R{\'e}ka M. and {M{\"o}stl}, Christian and {Lugaz}, No{\'e}},
        title = "{A Catalog of Interplanetary Coronal Mass Ejections Observed by Juno between 1 and 5.4 au}",
      journal = {\apj},
         year = 2021,
        month = dec,
       volume = {923},
       number = {2},
          eid = {136},
        pages = {136},
          doi = {10.3847/1538-4357/ac2ccb},
archivePrefix = {arXiv},
       eprint = {2111.11336},
 primaryClass = {physics.space-ph},
       adsurl = {https://ui.adsabs.harvard.edu/abs/2021ApJ...923..136D}
}

@ARTICLE{laker2024,
       author = {{Laker}, R. and {Horbury}, T.~S. and {O'Brien}, H. and {Fauchon-Jones}, E.~J. and {Angelini}, V. and {Fargette}, N. and {Amerstorfer}, T. and {Bauer}, M. and {M{\"o}stl}, C. and {Davies}, E.~E. and {Davies}, J.~A. and {Harrison}, R. and {Barnes}, D. and {Dumbovi{\'c}}, M.},
        title = "{Using Solar Orbiter as an Upstream Solar Wind Monitor for Real Time Space Weather Predictions}",
      journal = {Space Weather},
         year = 2024,
        month = feb,
       volume = {22},
       number = {2},
          eid = {e2023SW003628},
        pages = {e2023SW003628},
          doi = {10.1029/2023SW003628},
archivePrefix = {arXiv},
       eprint = {2307.01083},
 primaryClass = {physics.space-ph},
       adsurl = {https://ui.adsabs.harvard.edu/abs/2024SpWea..2203628L}
}

@ARTICLE{regnault2023,
       author = {{Regnault}, F. and {Al-Haddad}, N. and {Lugaz}, N. and {Farrugia}, C.~J. and {Yu}, W. and {Davies}, E.~E. and {Galvin}, A.~B. and {Zhuang}, B.},
        title = "{Investigating the Magnetic Structure of Interplanetary Coronal Mass Ejections Using Simultaneous Multispacecraft In Situ Measurements}",
      journal = {\apj},
         year = 2023,
        month = nov,
       volume = {957},
       number = {1},
          eid = {49},
        pages = {49},
          doi = {10.3847/1538-4357/acef16},
archivePrefix = {arXiv},
       eprint = {2309.10582},
 primaryClass = {astro-ph.SR},
       adsurl = {https://ui.adsabs.harvard.edu/abs/2023ApJ...957...49R}
}

@ARTICLE{moestl2022,
       author = {{M{\"o}stl}, Christian and {Weiss}, Andreas J. and {Reiss}, Martin A. and {Amerstorfer}, Tanja and {Bailey}, Rachel L. and {Hinterreiter}, J{\"u}rgen and {Bauer}, Maike and {Barnes}, David and {Davies}, Jackie A. and {Harrison}, Richard A. and {Freiherr von Forstner}, Johan L. and {Davies}, Emma E. and {Heyner}, Daniel and {Horbury}, Tim and {Bale}, Stuart D.},
        title = "{Multipoint Interplanetary Coronal Mass Ejections Observed with Solar Orbiter, BepiColombo, Parker Solar Probe, Wind, and STEREO-A}",
      journal = {\apjl},
         year = 2022,
        month = jan,
       volume = {924},
       number = {1},
          eid = {L6},
        pages = {L6},
          doi = {10.3847/2041-8213/ac42d0},
archivePrefix = {arXiv},
       eprint = {2109.07200},
 primaryClass = {astro-ph.SR},
       adsurl = {https://ui.adsabs.harvard.edu/abs/2022ApJ...924L...6M}
}

@ARTICLE{salman_2024_psp_cat,
       author = {{Salman}, Tarik Mohammad and {Nieves-Chinchilla}, Teresa and {Jian}, Lan K. and {Lugaz}, No{\'e} and {Carcaboso}, Fernando and {Davies}, Emma E. and {Collado-Vega}, Yaireska M.},
        title = "{A Survey of Coronal Mass Ejections Measured In Situ by Parker Solar Probe during 2018{\textendash}2022}",
      journal = {\apj},
         year = 2024,
        month = may,
       volume = {966},
       number = {1},
          eid = {118},
        pages = {118},
          doi = {10.3847/1538-4357/ad320c},
archivePrefix = {arXiv},
       eprint = {2403.02594},
 primaryClass = {physics.space-ph},
       adsurl = {https://ui.adsabs.harvard.edu/abs/2024ApJ...966..118S}
}

@ARTICLE{luhmann_2020_review,
       author = {{Luhmann}, J.~G. and {Gopalswamy}, N. and {Jian}, L.~K. and {Lugaz}, N.},
        title = "{ICME Evolution in the Inner Heliosphere}",
      journal = {\solphys},
         year = 2020,
        month = apr,
       volume = {295},
       number = {4},
          eid = {61},
        pages = {61},
          doi = {10.1007/s11207-020-01624-0},
       adsurl = {https://ui.adsabs.harvard.edu/abs/2020SoPh..295...61L}
}

@ARTICLE{Davies_2021_solo,
       author = {{Davies}, E.~E. and {M{\"o}stl}, C. and {Owens}, M.~J. and {Weiss}, A.~J. and {Amerstorfer}, T. and {Hinterreiter}, J. and {Bauer}, M. and {Bailey}, R.~L. and {Reiss}, M.~A. and {Forsyth}, R.~J. and {Horbury}, T.~S. and {O'Brien}, H. and {Evans}, V. and {Angelini}, V. and {Heyner}, D. and {Richter}, I. and {Auster}, H. -U. and {Magnes}, W. and {Baumjohann}, W. and {Fischer}, D. and {Barnes}, D. and {Davies}, J.~A. and {Harrison}, R.~A.},
        title = "{In situ multi-spacecraft and remote imaging observations of the first CME detected by Solar Orbiter and BepiColombo}",
      journal = {\aap},
         year = 2021,
        month = dec,
       volume = {656},
          eid = {A2},
        pages = {A2},
          doi = {10.1051/0004-6361/202040113},
archivePrefix = {arXiv},
       eprint = {2012.07456},
 primaryClass = {physics.space-ph},
       adsurl = {https://ui.adsabs.harvard.edu/abs/2021A&A...656A...2D}
}

@ARTICLE{moestl2020,
       author = {{M{\"o}stl}, Christian and {Weiss}, Andreas J. and {Bailey}, Rachel L. and {Reiss}, Martin A. and {Amerstorfer}, Tanja and {Hinterreiter}, J{\"u}rgen and {Bauer}, Maike and {McIntosh}, Scott W. and {Lugaz}, No{\'e} and {Stansby}, David},
        title = "{Prediction of the In Situ Coronal Mass Ejection Rate for Solar Cycle 25: Implications for Parker Solar Probe In Situ Observations}",
      journal = {\apj},
         year = 2020,
        month = nov,
       volume = {903},
       number = {2},
          eid = {92},
        pages = {92},
          doi = {10.3847/1538-4357/abb9a1},
archivePrefix = {arXiv},
       eprint = {2007.14743},
 primaryClass = {astro-ph.SR},
       adsurl = {https://ui.adsabs.harvard.edu/abs/2020ApJ...903...92M}
}

@ARTICLE{leitner_2007,
       author = {{Leitner}, M. and {Farrugia}, C.~J. and {M{\"o}Stl}, C. and
         {Ogilvie}, K.~W. and {Galvin}, A.~B. and {Schwenn}, R. and
         {Biernat}, H.~K.},
        title = "{Consequences of the force-free model of magnetic clouds for their heliospheric evolution}",
      journal = {Journal of Geophysical Research (Space Physics)},
         year = "2007",
        month = "Jun",
       volume = {112},
       number = {A6},
          eid = {A06113},
        pages = {A06113},
          doi = {10.1029/2006JA011940},
       adsurl = {https://ui.adsabs.harvard.edu/abs/2007JGRA..112.6113L}
}

@ARTICLE{bothmer1998,
   author = {{Bothmer}, V. and {Schwenn}, R.},
    title = "{The structure and origin of magnetic clouds in the solar wind}",
  journal = {\angeo},
     year = 1998,
    month = jan,
   volume = 16,
    pages = {1-24},
      doi = {10.1007/s00585-997-0001-x},
   adsurl = {http://adsabs.harvard.edu/abs/1998AnGeo..16....1B}
}

@ARTICLE{burlaga1981,
   author = {{Burlaga}, L. and {Sittler}, E. and {Mariani}, F. and {Schwenn}, R.
	},
    title = "{Magnetic loop behind an interplanetary shock - Voyager, Helios, and IMP 8 observations}",
  journal = {\jgr},
     year = 1981,
    month = aug,
   volume = 86,
    pages = {6673-6684},
      doi = {10.1029/JA086iA08p06673},
   adsurl = {http://adsabs.harvard.edu/abs/1981JGR....86.6673B}
}

@ARTICLE{cyr2000,
       author = {{Cyr}, O.~C.~S. and {Mesarch}, M.~A. and {Maldonado}, H.~M. and
         {Folta}, D.~C. and {Harper}, A.~D. and {Davila}, J.~M. and
         {Fisher}, R.~R.},
        title = "{Space Weather Diamond: a four spacecraft monitoring system}",
      journal = {\jastp},
         year = "2000",
        month = "Sep",
       volume = {62},
       number = {14},
        pages = {1251-1255},
          doi = {10.1016/S1364-6826(00)00069-9},
       adsurl = {https://ui.adsabs.harvard.edu/abs/2000JASTP..62.1251C}
}

@ARTICLE{good2018,
   author = {{Good}, S.~W. and {Forsyth}, R.~J. and {Eastwood}, J.~P. and 
	{M{\"o}stl}, C.},
    title = "{Correlation of ICME Magnetic Fields at Radially Aligned Spacecraft}",
  journal = {\solphys},
archivePrefix = "arXiv",
   eprint = {1802.04004},
 primaryClass = "physics.space-ph",
     year = 2018,
    month = mar,
   volume = 293,
      eid = {52},
    pages = {52},
      doi = {10.1007/s11207-018-1264-y},
   adsurl = {http://adsabs.harvard.edu/abs/2018SoPh..293...52G}
}

@ARTICLE{kubicka2016,
   author = {{Kubicka}, M. and {M{\"o}stl}, C. and {Amerstorfer}, T. and 
	{Boakes}, P.~D. and {Feng}, L. and {Eastwood}, J.~P. and {T{\"o}rm{\"a}nen}, O.
	},
    title = "{Prediction of Geomagnetic Storm Strength from Inner Heliospheric In Situ Observations}",
  journal = {\apj},
archivePrefix = "arXiv",
   eprint = {1610.06713},
 primaryClass = "physics.space-ph",
     year = 2016,
    month = dec,
   volume = 833,
      eid = {255},
    pages = {255},
      doi = {10.3847/1538-4357/833/2/255},
   adsurl = {http://adsabs.harvard.edu/abs/2016ApJ...833..255K}
}

@ARTICLE{lepping1990,
   author = {{Lepping}, R.~P. and {Burlaga}, L.~F. and {Jones}, J.~A.},
    title = "{Magnetic field structure of interplanetary magnetic clouds at 1 AU}",
  journal = {\jgr},
     year = 1990,
    month = aug,
   volume = 95,
    pages = {11957-11965},
      doi = {10.1029/JA095iA08p11957},
   adsurl = {http://adsabs.harvard.edu/abs/1990JGR....9511957L}
}

@ARTICLE{lindsay1999,
       author = {{Lindsay}, G.~M. and {Russell}, C.~T. and {Luhmann}, J.~G.},
        title = "{Predictability of Dst index based upon solar wind conditions monitored inside 1 AU}",
      journal = {\jgr},
         year = "1999",
        month = "May",
       volume = {104},
       number = {A5},
        pages = {10335-10344},
          doi = {10.1029/1999JA900010},
       adsurl = {https://ui.adsabs.harvard.edu/abs/1999JGR...10410335L}
}

@ARTICLE{liu2014,
       author = {{Liu}, Ying D. and {Luhmann}, Janet G. and {Kajdi{\v{c}}}, Primo{\v{z}}
        and {Kilpua}, Emilia K.~J. and {Lugaz}, No{\'e} and {Nitta},
        Nariaki V. and {M{\"o}stl}, Christian and {Lavraud}, Benoit and
        {Bale}, Stuart D. and {Farrugia}, Charles J. and {Galvin},
        Antoinette B.},
        title = "{Observations of an extreme storm in interplanetary space caused by
        successive coronal mass ejections}",
      journal = {\natco},
         year = 2014,
        month = Mar,
       volume = {5},
          eid = {3481},
        pages = {3481},
          doi = {10.1038/ncomms4481},
archivePrefix = {arXiv},
       eprint = {1405.6088},
 primaryClass = {physics.space-ph},
       adsurl = {https://ui.adsabs.harvard.edu/\#abs/2014NatCo...5E3481L}
}

@ARTICLE{manchester2017,
   author = {{Manchester}, W. and {Kilpua}, E.~K.~J. and {Liu}, Y.~D. and 
	{Lugaz}, N. and {Riley}, P. and {T{\"o}r{\"o}k}, T. and {Vr{\v s}nak}, B.
	},
    title = "{The Physical Processes of CME/ICME Evolution}",
  journal = {\ssr},
     year = 2017,
    month = aug,
      doi = {10.1007/s11214-017-0394-0},
   adsurl = {http://adsabs.harvard.edu/abs/2017SSRv..tmp...90M}
}

@ARTICLE{mulligan1998,
   author = {{Mulligan}, T. and {Russell}, C.~T. and {Luhmann}, J.~G.},
    title = "{Solar cycle evolution of the structure of magnetic clouds in the inner heliosphere}",
  journal = {\grl},
     year = 1998,
   volume = 25,
    pages = {2959-2962},
      doi = {10.1029/98GL01302},
   adsurl = {http://adsabs.harvard.edu/abs/1998GeoRL..25.2959M}
}

@ARTICLE{owens2017,
   author = {{Owens}, M.~J. and {Lockwood}, M. and {Barnard}, L.~A.},
    title = "{Coronal mass ejections are not coherent magnetohydrodynamic structures}",
  journal = {\natsr},
     year = 2017,
    month = jun,
   volume = 7,
      eid = {4152},
    pages = {4152},
      doi = {10.1038/s41598-017-04546-3},
   adsurl = {http://adsabs.harvard.edu/abs/2017NatSR...7.4152O}
}
\bibliographystyle{aasjournalv7}


\end{document}